\documentclass[twocolumn]{aastex631}
\usepackage[percent]{overpic}
\usepackage{graphicx}

\begin{document}

%\title{COSMIC:A search for narrow-band technosignatures toward planet K2-18b}
%\title{A Multi-Facility Search for Narrowband Technosignatures Toward K2–18b}
\title{A Narrowband Technosignature Search Toward the Hycean Candidate K2-18b Using the VLA and MeerKAT}

\author[0000-0002-4409-3515]{Tremblay, C.D.}
\affiliation{SETI Institute, 339 Bernardo Ave, Suite 200, Mountain View, CA 94043, USA}
\affiliation{Berkeley SETI Research Center, University of California, Berkeley, CA 94720, USA}
\affiliation{Department of Physics and Astronomy, University of New Mexico, Albuquerque, NM 87131, USA}
\affiliation{National Radio Astronomy Observatory, 1003 Lopezville Rd., Socorro, NM 87801, USA}

\author{Chaudhary, S.}
\affiliation{Indian Institute of Science Education and Research, Mohali, India}

\author[0000-0002-3012-4261]{Li, Megan G.}
\affiliation{Department of Earth, Planetary, and Space Sciences, University of California, Los Angeles, CA 90024, USA}

\author[0000-0001-7057-4999]{Sheikh, Sofia Z.}
\affiliation{SETI Institute, 339 Bernardo Ave, Suite 200, Mountain View, CA 94043, USA}
\affiliation{Berkeley SETI Research Center, University of California, Berkeley, CA 94720, USA}
\author[0000-0003-0804-9362]{T. Myburgh}
\affiliation{Mydon Solutions (Pty) Ltd., 102 Silver Oaks, 23 Silverlea Road, Wynberg, Cape Town, South Africa, 7800}
\affiliation{SETI Institute, 339 Bernardo Ave, Suite 200, Mountain View, CA 94043, USA}
\author[0000-0002-8071-6011]{D. Czech}
\affiliation{Berkeley SETI Research Center, University of California, Berkeley, CA 94720, USA}
\author[0000-0001-6950-5072]{D.E. MacMahon }
\affiliation{Berkeley SETI Research Center, University of California, Berkeley, CA 94720, USA}
\author[0000-0002-6664-965X]{P.B. Demorest}
\affiliation{National Radio Astronomy Observatory, 1003 Lopezville Rd., Socorro, NM 87801, USA}
\author[0009-0001-8677-372X]{R.A. Donnachie}
\affiliation{Mydon Solutions (Pty) Ltd., 102 Silver Oaks, 23 Silverlea Road, Wynberg, Cape Town, South Africa, 7800}
\affiliation{SETI Institute, 339 Bernardo Ave, Suite 200, Mountain View, CA 94043, USA}
\author[0000-0003-2828-7720]{A.P.V. Siemion }
\affiliation{SETI Institute, 339 Bernardo Ave, Suite 200, Mountain View, CA 94043, USA}
\affiliation{Breakthrough Listen, University of Oxford, Department of Physics, Denys Wilkinson Building, Keble Road, Oxford, OX1 3RH, UK}
\author[0000-0002-8604-106X]{Gajjar, V.}
\affiliation{SETI Institute, 339 Bernardo Ave, Suite 200, Mountain View, CA 94043, USA}
\author[0000-0002-7042-7566]{Lebofsky, M.}
\affiliation{Berkeley SETI Research Center, University of California, Berkeley, CA 94720, USA}
\author[0000-0003-4338-2611]{Wandia, K.}
\affiliation{Jodrell Bank Centre for Astrophysics, University of Manchester, M13 9PL, UK}
\author[0000-0002-6341-4548]{~Perez, K I.}
\affiliation{SETI Institute, 339 Bernardo Ave, Suite 200, Mountain View, CA 94043, USA}
%\author{Others add as contribute}
\author[0000-0002-4869-000X]{Nikku Madhusudhan}
\affiliation{Institute of Astronomy, University of Cambridge, Madingley Road, Cambridge CB3 0HA, UK}



\begin{abstract}
K2--18b, a sub-Neptune exoplanet located in the habitable zone of its host star, has emerged as an important target for atmospheric characterization, and assessments of potential habitability. Motivated by recent interpretations of \textit{JWST} observations suggesting a hydrogen-rich atmosphere consistent with Hycean-world scenarios, we conducted a coordinated, multi-epoch search for narrowband radio technosignatures using the Karl G.\ Jansky Very Large Array equipped with the \textsc{COSMIC} backend and the MeerKAT telescope with the \textsc{BLUSE} backend. Our observations span frequencies from 544\,MHz to 9.8\,GHz and include multiple epochs that cover at least one full orbital period of the planet.  In this work we outline, create, and apply a comprehensive post-processing framework that incorporates observatory-informed RFI masking, drift-rate filtering based on the expected dynamics of the K2--18 system, multibeam spatial discrimination, primary and secondary transit filtering (when applicable), and SNR-based excision of weak and strong spurious signals. Across all bands and epochs, no signals consistent with an astrophysical or artificial origin were identified at a limit of 10$^{12}$ to 10$^{13}$\,W. These non-detections allow us to place upper limits on the presence of persistent, isotropic narrowband transmitters within the K2--18 system, providing the first interferometric technosignature constraints for a Hycean-planet candidate. Our results demonstrate the efficacy of coordinated multi-epoch interferometric searches and establish a methodological framework for future technosignature studies of nearby potentially habitable exoplanets.
\end{abstract}



\section{Introduction} \label{sec:intro}
K2-18b, a sub-Neptune exoplanet located within the habitable zone of its host star, presents a compelling opportunity to investigate the atmospheric composition and potential habitability of planets beyond our Solar System. With an estimated mass of 8.63 $\pm$ 1.35\,$\rm{M}_\odot$, a radius of 2.61 $\pm$ 0.09\,$\rm{R}_\odot$, and an equilibrium temperature between 250--300\,K \citep{Montet_2015, sarkis2018, cloutier2019, Benneke2019}, K2-18b may possess conditions suitable for liquid water \citep{Madhusudhan_2020}. Atmospheric studies using transmission spectra from the James Webb Space Telescope (JWST) indicate the presence of substantial CH$_4$ and CO$_2$, while showing a lack of NH$_3$ and stratospheric H$_2$O \citep{Madhusudhan_2023}. These findings are consistent with theoretical models predictions \citep[e.g.][]{Hu_2021, Tsai_2021} for Hycean worlds — planets characterized by deep, stable oceans beneath hydrogen-rich atmospheres \citep{Madhusudhan_2021}. Such a chemical composition raises intriguing questions regarding the planet’s potential habitability and the persistence of surface oceans \citep{Barrier_2025, Hu_2025}.

%Observations from the Hubble Space Telescope (HST) have introduced additional complexity, suggesting that apparent atmospheric features, particularly those attributed to water vapor, may be affected by stellar heterogeneities such as starspots \citep{Barclay_2021}. To evaluate this possibility, \cite{Barclay_2021} modeled the variability of the host star K2-18 based on its photometric light curve, generating 1,000 synthetic HST datasets replicating the K2-18b observing strategy. Approximately 1\% of these simulations produced better fits to the data than the exoplanet atmospheric model, and 40 \% yielded water detections of comparable significance. These findings imply that stellar contamination may influence the inferred atmospheric composition, emphasizing the need for refined models and higher-precision observations to accurately characterize K2-18b’s atmosphere.

Thus, K2-18b serves as an important benchmark for studies of sub-Neptune atmospheres and their potential habitability. At the same time, it highlights the importance of cautious interpretation when analyzing transmission spectra of planets orbiting magnetically active stars \citep[e.g][]{Barclay_2021}. Ongoing and future observations with JWST and complementary facilities are expected to further constrain the planet’s atmospheric structure, composition, and dynamics, providing key insights into Hycean worlds and the broader search for habitable environments beyond Earth and potential hosts for atmospheric biosignatures.

Although the potential for biosignatures in K2-18b’s atmosphere remains tentative \citep[e.g.][]{Madhusudhan_2023, Madhusudhan_2025, Welbanks_2025, Pica-ciamarra_2025}, another way to constrain the planet's habitability is to consider the search for technosignatures---observable indicators of advanced technology within or near the planetary system. The leakage radiation from the Earth consists of a large fraction of narrowband signals ($<<$500\,Hz) and, in addition, narrowband signals are ideal beacons \citep{Cocconi_1959,Wright_2022,Sheikh_2025}.
%The detection of electromagnetic radiation consistent with artificial emission would offer a distinct line of evidence for technological activity. Broadband emission ($>$ 0.5 kHz) would be difficult to distinguish from natural astrophysical or chemical processes, but narrowband signals (a few Hz wide) originating from the K2-18 system would constitute a more definitive indicator of artificial origin \citep{Wright_2022}. 
Additionally, since almost 80\% of Earth’s leakage from powerful radar signals such as the Deep Space Network (DSN) is located in the ecliptic plane \citep{Fan_2025}, it is a high probability that when looking at the ecliptic planes of other systems, the chance of detection of a technosignature increases. To distinguish genuine extraterrestrial signals from terrestrial radio frequency interference (RFI), we leverage the planet’s orbital motion and stellar occultation events: for example, true signals from K2-18b would exhibit predictable Doppler drift rates and any emission should vanish when the planet passes behind its host star.

To explore this possibility, we have undertaken a coordinated observing campaign with the MeerKAT radio telescope in South Africa and the National Science Foundation’s Karl G. Jansky Very Large Array (VLA) in New Mexico. This observing program spans frequencies from 544\,MHz to 9.8\,GHz and multiple epochs, covering at least one full orbital period of K2-18b. This is in addition to the campaign completed at the Allen Telescope Array in California \citep{Saide_2023}. These observations aim to search for narrowband technosignatures while simultaneously advancing our understanding of the atmospheric and environmental properties of the planet. In addition to the observations, in this work we provide a framework for the post-processing of observations toward other exoplanetary systems.

\section{Observations} \label{sec:style}

\begin{deluxetable}{lcccc}
\tablecaption{Summary of Observations Toward K2-18b \label{tab:obs_summary}}
\tablehead{
\colhead{Telescope} & \colhead{Band} & \colhead{Obs. Date} & \colhead{Obs. Time} & \colhead{Orbital Phase\tablenotemark{a}}
\\
& & & (UTC) & ($\phi$)}
\startdata
VLA      & S-Band & 2023-10-03 & 19:50--20:00 & 0.76 \\
VLA      &  & 2023-10-05 & 20:28--20:38 & 0.82 \\
VLA      &  & 2023-10-13 & 15:03--15:13 & 0.06 \\
VLA      &  & 2023-10-22 & 14:26--14:36 & 0.33 \\
VLA      &  & 2023-12-08 & 17:31--17:41 & 0.76 \\
VLA      &  & 2023-12-14 & 10:49--10:59 & 0.94 \\
VLA      &  & 2023-12-21 & 08:29--08:39 & 0.15 \\
VLA      & C-Band & 2023-12-08 & 17:01--17:11 & 0.76 \\
VLA      &  & 2023-12-14 & 10:18--10:28 & 0.94 \\
VLA      &  & 2023-12-19 & 15:44--15:54 & 0.09 \\
VLA      & X-Band & 2023-12-08 & 10:02--10:12 & 0.75 \\
VLA      &  & 2023-12-13 & 09:25--09:35 & 0.90 \\
VLA      &  & 2023-12-19 & 16:43--16:53 & 0.10 \\
MeerKAT  & UHF    & 2023-09-30 & 11:13--11:18 & 0.66 \\
MeerKAT  & L-Band & 2023-10-01 & 11:03--11:08 & 0.69 \\
MeerKAT  & S4-Band& 2023-10-06 & 08:48--08:53 & 0.84 \\
\enddata
\tablenotetext{a}{Calculated from \citet{Benneke2019}, where T0=57725.051189 MJD (UTC) for the mid-transit ($\phi$=0).}
\end{deluxetable}

\subsection{VLA-COSMIC}
\begin{figure*}
    \centering
    \includegraphics[width=0.48\linewidth]{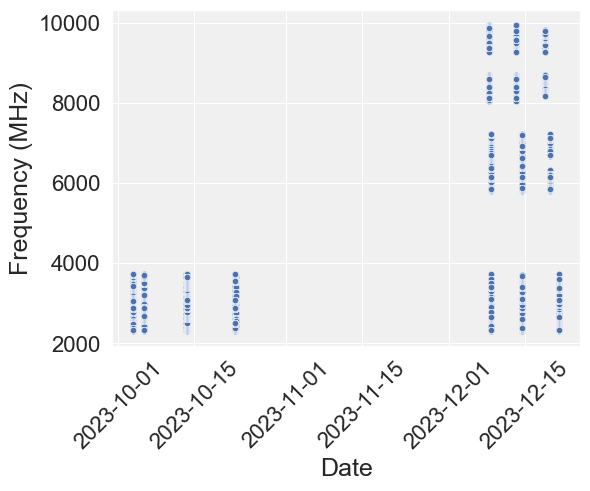}
    \includegraphics[width=0.42\linewidth]{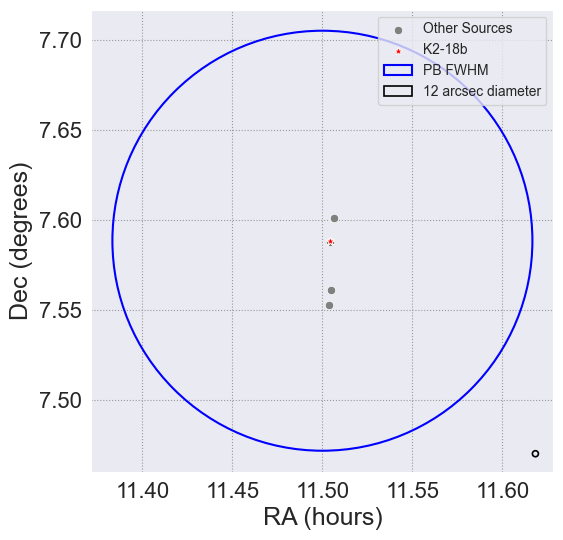}
    \caption{The observed sources and data collected during the K2-18b observing campaign on COSMIC. (Left) The dates of the observations and the frequencies in which signals were detected. (Right) The location of the 4 coherent beams (gray circles), the position of K2-18 (red star), and the full-width-half-maximum primary beam (FWHM PB) for the COSMIC telescope at S-band.}
    \label{fig:COSMICdata}
\end{figure*}

We observed the K2-18 planetary system and its host star using the VLA. Observations were conducted over multiple epochs, spaced up to one week apart, between 29 September 2023 and 21 December 2023 (Project Code: VLA/23B-307), as shown in Figure \ref{fig:COSMICdata}. The dates of the observations for each band are summarized in Table \ref{tab:obs_summary} and the parameters for the real-time processing are outlined in Table \ref{tab:obs_setup}.

Data were collected simultaneously using both the Commensal Open-Source Multi-Mode Interferometer Cluster (COSMIC) backend \citep{tremblay_cosmic} and the standard $WIDAR$ correlator system \citep{perley_widar}. For the standard $WIDAR$ data, the correlator was configured in 8-bit mode with a 2\,GHz bandwidth and operated in continuum mode with 2\,MHz-wide coarse channels. These data were analyzed for radio emission from the host; no detections were made for either the coherent or incoherent emission \citep{Wandia_2025}.

The \textsc{COSMIC} system is designed to operate in parallel with most VLA observations, recording data passively without direct control over the observing frequency or pointing. For this program, however, we were awarded six hours of Director’s Discretionary Time (Proposal VLA/23B-307). During these observations, we recorded up to 1.2\,GHz of bandwidth across three frequency ranges: 2.4–3.6\,GHz (\textit{S}-band), 5--7.2\,GHz (\textit{C}-band), and 8.1–9.8\,GHz (\textit{X}-band). Over the course of the campaign, the source was observed seven times in \textit{S}-band and three times each in \textit{C}- and \textit{X}-band. For each epoch, the source was observed for a total of 10 minutes, with each 56\,second segment independently processed and searched for narrowband signals. See Figure \ref{fig:COSMICdata} for an outline of the dates and frequency coverage of the observations, and the placement of the 4 coherent beams in the field.

\subsubsection{Calibration and real-time processing}
The \textsc{COSMIC} system employs a real-time data-processing pipeline that performs calibration, channelization, beamforming, and searches for narrowband signals. A detailed description of this pipeline is provided in \citet{tremblay_cosmic} and \citet{Tremblay_VLASS}; a brief summary is given here. The pipeline is controlled through an observing script that defines the triggers for the experiment (e.g., observation ID, project code, and frequency range) and instructs the system when to record and how to process the data. When an observation is triggered, the VLA software assigns an observation intent\footnote{A full list of the observations intents are noted on the NRAO website. \url{https://www.aoc.nrao.edu/~cwalker/sched/INTENTs.html}} that initiates either a correlation and calibration sequence or a technosignature search mode.

During the calibration phase, the data are divided into 32 fragments of 1\,MHz each and distributed across processing nodes, where correlation is performed using the \textsc{xGPU} software correlator. Complex gain solutions are then derived for each frequency segment. The results are returned to the head node, where the data are reordered by frequency and a calibration quality assessment is completed. Real-time RFI flagging is also performed, and gain solutions are smoothed over flagged 1\,MHz channels. The final gain solutions are applied to the field-programmable gate arrays (FPGAs), enabling automatic phase calibration of incoming target-field data.

\subsection{MeerKAT-BLUSE}
We conducted single-epoch observations in each frequency band toward K2-18 using the MeerKAT telescope in South Africa with the commensal backend, \textsc{Breakthrough Listen User Supplied Equipment} (\textsc{BLUSE}; \citealt{Czech_2021}, Czech et al., in prep.). The observations were performed on 2023 September 30 at \textit{UHF} (544-1015\,MHz), 2023 October 1 at \textit{L}-band (900-1670\,MHz), and 2023 October 6 at \textit{S4} (2625–3500\,MHz). A summary of the observations and dates are listed in Table \ref{tab:obs_summary} and the parameters for the real-time processing are outlined in Table \ref{tab:obs_setup}.. 

BLUSE is an autonomous commensal SETI survey system. It ingests and processes packetized raw voltage data (coarse channels) directly from MeerKAT's F-engines via Ethernet multicast groups. Ordinarily, it operates in an autonomous commensal fashion, configuring itself to process the incoming data for any observing mode selected by the primary observer. It selects up to 64 targets of interest that happen to fall within the primary field of view, channelizes the data to $\sim$1\,Hz resolution, forms synthesized beams on the targets, and conducts a narrowband technosignature search for each beam (seticore\footnote{\url{https://github.com/lacker/seticore}}, similar to COSMIC).

For this experiment, however, BLUSE was allocated primary telescope time (Director's Discretionary Time Proposal ID: DDT-20230920-DC-01) in a "semi-commensal" mode, an approach designed to use telescope time efficiently by leveraging BLUSE's ability to operate alongside any configuration selected by a primary observer. BLUSE can record up to 290 seconds worth of MeerKAT's full bandwidth before its buffers are full and it needs to spend time processing the data. Therefore, a small amount of primary BLUSE telescope time was appended to the beginning of several primary observations: MeerKAT was configured and the data calibrated for the primary observation, but was directed to observe K2-18 for 5 minutes prior to the start of the primary observation.  

\begin{figure*}
    \centering
    \includegraphics[width=0.54\linewidth]{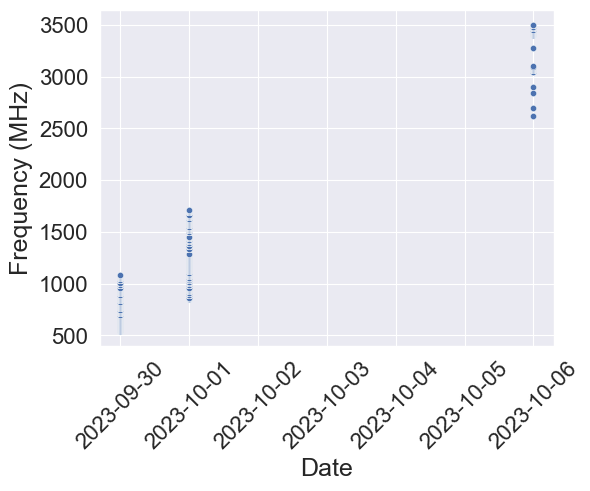}
    \includegraphics[width=0.45\linewidth]{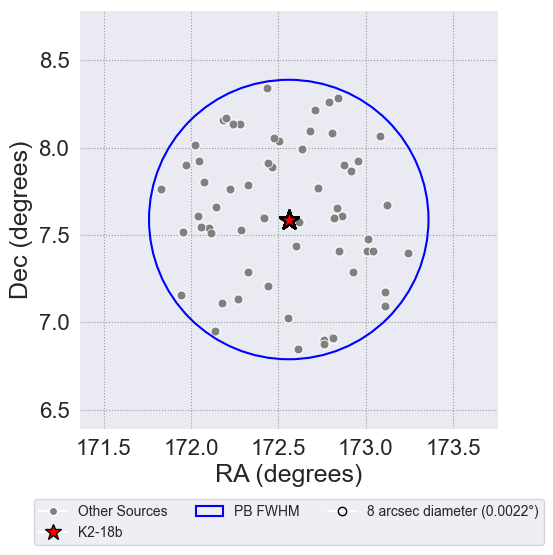}
    \caption{The observed sources and data collected during the K2-18b observing campaign. (Left) The dates of the observations and the frequencies in which signals were detected. (Right) The location of the 64 coherent beams (gray circles), the position of K2-18 (red star), and the FWHM PB for the MeerKAT telescope at L-band.}
    \label{fig:MeerKATdata}
\end{figure*}

This approach allowed BLUSE to record, process, and store additional data products (which take considerably longer to write to disk but allow later flexibility during analysis) at leisure without occupying further primary telescope time. Since BLUSE is able to operate alongside all primary telescope modes, it was easy to find suitable primary observations in UHF, L, and S-band. See Figure \ref{fig:MeerKATdata} for the observations and the layout of the 64 coherent beams within the telescope's field of view (FOV) during the L-band observation.

\begin{deluxetable*}{lcccccccc}
\tablecaption{Parameters for the COSMIC and BLUSE observations and real-time pipeline.\label{tab:obs_setup}}
\tablewidth{0pt}
\tablehead{
\colhead{Telescope} &
\colhead{Band} &
\colhead{Date} &
\colhead{Integration Time} &
\colhead{Threshold} &
\colhead{Channel Width} &
\colhead{No. of Antennas} &
\colhead{Recording*} &
\colhead{Polarizations} \\
&
&
&
\colhead{(s)} &
\colhead{($\sigma$)} &
\colhead{(Hz)} &
&
&
}
\startdata
VLA     & S  & 2023-10-03 & 56  & 10 & 2 & 17 & 8-bit  & 2 \\
VLA     & S  & 2023-10-05 & 56  & 10 & 2 & 23 & 8-bit  & 2 \\
VLA     & S  & 2023-10-13 & 56  & 10 & 2 & 24 & 8-bit  & 2 \\
VLA     & S  & 2023-10-22 & 56  & 10 & 2 & 22 & 8-bit  & 2 \\
VLA     & S  & 2023-12-08 & 56  & 10 & 2 & 25 & 8-bit  & 2 \\
VLA     & S  & 2023-12-14 & 56  & 10 & 2 & 25 & 8-bit  & 2 \\
VLA     & S  & 2023-12-21 & 56  & 10 & 2 & 23 & 8-bit  & 2 \\
VLA     & C  & 2023-12-08 & 56  & 10 & 2 & 25 & 8-bit  & 2 \\
VLA     & C  & 2023-12-14 & 56  & 10 & 2 & 25 & 8-bit  & 2 \\
VLA     & C  & 2023-12-19 & 56  & 10 & 2 & 25 & 8-bit  & 2 \\
VLA     & X  & 2023-12-08 & 56  & 10 & 2 & 25 & 8-bit  & 2 \\
VLA     & X  & 2023-12-13 & 56  & 10 & 2 & 25 & 8-bit  & 2 \\
VLA     & X  & 2023-12-19 & 56  & 10 & 2 & 25 & 8-bit  & 2 \\
MeerKAT & UHF & 2023-09-30 & 290 & 10 & 1 & 64 & 10-bit & 2 \\
MeerKAT & L   & 2023-10-01 & 290 & 10 & 1 & 64 & 10-bit & 2 \\
MeerKAT & S4  & 2023-10-06 & 290 & 10 & 1 & 64 & 10-bit & 2 \\
\enddata
\tablenotetext{}{*For both 8-bit and 10-bit recordings, quantization losses are negligible for the sensitivity limits considered in this work.}
\end{deluxetable*}

\subsection{Technosignature Candidate Search}
To search for narrowband signals, the pipeline fine-channelized the data from the coarse channels into fine-frequency bins using a Fast Fourier Transform (FFT)---2\,Hz for \textsc{COSMIC} and $\sim$1\,Hz for \textsc{BLUSE}\footnote{The exact frequency resolution changes depending on primary configuration.}. The \textsc{COSMIC} system forms up to four coherent beams and an incoherent sum across the field of view (FOV), while MeerKAT forms 64 coherent beams. The VLA observations for this project were conducted during the array’s transition from the A configuration (36\,km baselines) to the D configuration (1\,km baselines); consequently, the synthesized beam size varied between epochs. Given the planet’s radius of 2.34\,R$_{\oplus}$ and a distance of 38\,pc, the entire K2-18 planetary system, including its host star, lies well within a single coherent beam of both telescopes.

For both facilities, a Taylor-tree de-dispersion algorithm implemented in the \textsc{seticore} software package was used to identify narrowband drifting signals within the real-time pipelines. The output of the search consists of a series of metadata entries stored in an SQL database, along with calibrated voltage files for each antenna centered on the detected signals (hits). The resulting data products were then filtered to identify candidate technosignatures originating from the K2-18 system.

\section{Post-Processing Methodology}
\label{PPM}
\begin{figure*}
    \centering
    \includegraphics[width=0.8\linewidth]{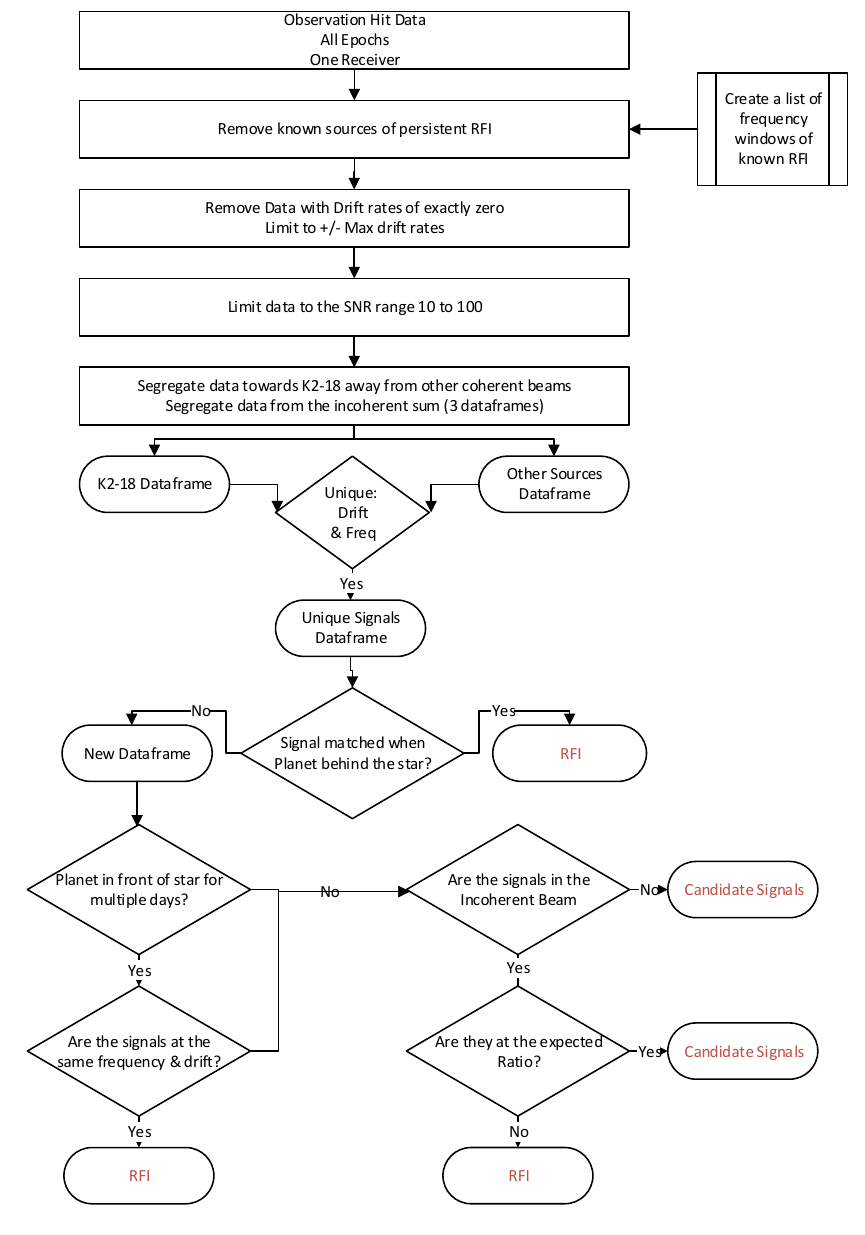}
    \caption{A block diagram summarizing the steps listed in Section 3 of the paper to filter the RFI discovered by the pipeline to find potential technosignature candidates.}
    \label{fig:pipeline}
\end{figure*}

The primary goal of this project is to search for evidence of technological activity within the K2-18 planetary system. We assess this possibility by examining whether signals detected by the real-time pipeline are spatially coincident with the planet’s position on the sky, exhibit the expected Doppler drift rates, and vanish when the planet passes behind its host star during secondary transit. %Based on prior work, we also impose limits on the signal-to-noise ratio (SNR), anticipating that any genuine technosignature would be weak, with an SNR below 100.\footnote{We note that this may not be the case for all technosignatures. 
%However, this is done to remove known systematics and the data are still present in the database and can be evaluated again at a later date with improved algorithms to remove these systematics.} 
%Experience from previous analyses indicates that the brightest signals are typically attributable to local radio-frequency interference (RFI) or instrumental systematics where the signal is limited to a single antenna and not across the array as expected for an astronomical source. 
The following section describes in detail the procedure used to filter the signals detected by the real-time pipelines to identify the most promising technosignature candidates and is summarized in Figure \ref{fig:pipeline}

\subsection{Remove Known Sources of RFI}
In this work, we employ multiple techniques to identify and mitigate sources of persistent RFI. For most of the frequency bands observed with both telescopes, we rely on observatory-provided data products. The NRAO supplies updated RFI plots for each configuration and observing semester, derived from time-averaged spectra in 128\,MHz partitions\footnote{\url{https://science.nrao.edu/facilities/vla/docs/manuals/obsguide/rfi}}. In addition, a table of known strong RFI sources is provided, enabling us to exclude channels containing the most powerful emitters.

The South African Radio Astronomy Observatory (SARAO) provides similar information for the MeerKAT telescope\footnote{\url{https://skaafrica.atlassian.net/wiki/spaces/ESDKB/pages/305332225/Radio+Frequency+Interference+RFI}}, though regularly updated reference plots are not yet available.

\begin{figure*}
    \centering
    \includegraphics[width=0.78\linewidth]{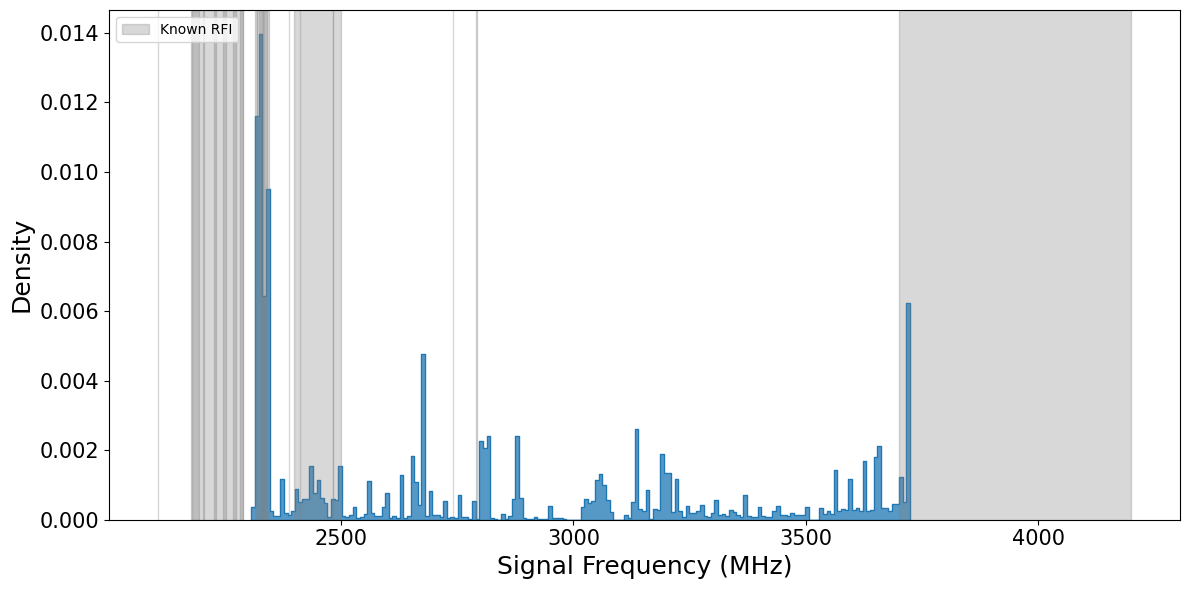}
    \includegraphics[width=0.78\linewidth]{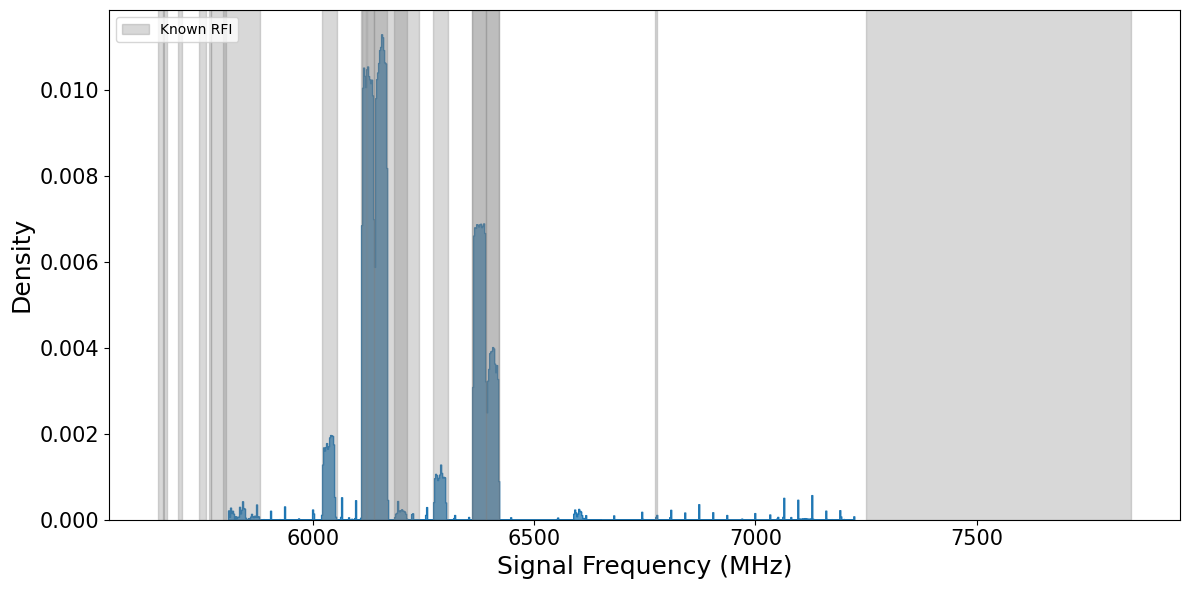}
    \includegraphics[width=0.78\linewidth]{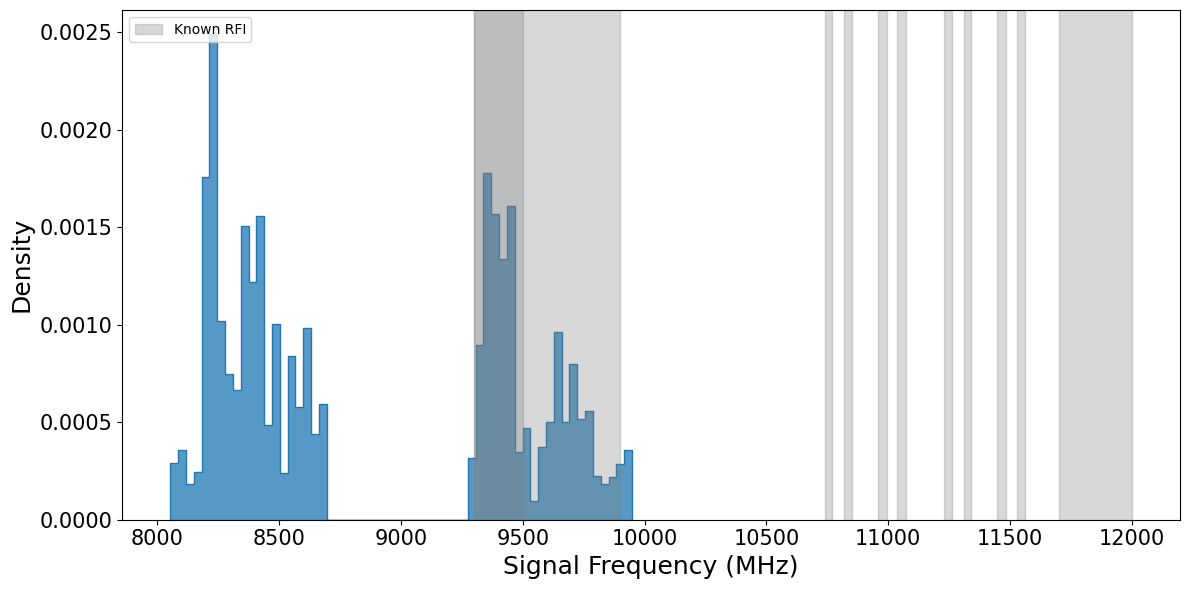}
    \caption{Histogram density plots of all the signals detected through the autonomous real-time pipeline running on COSMIC at the VLA prior to RFI rejection. The plots show the data from S-band (Top), C-band (middle), and X-band (bottom) representing all signals from all of the epochs of observations.}
    \label{fig:c-band-RFI}
\end{figure*}

\begin{figure*}
    \centering
    \includegraphics[width=0.95\linewidth]{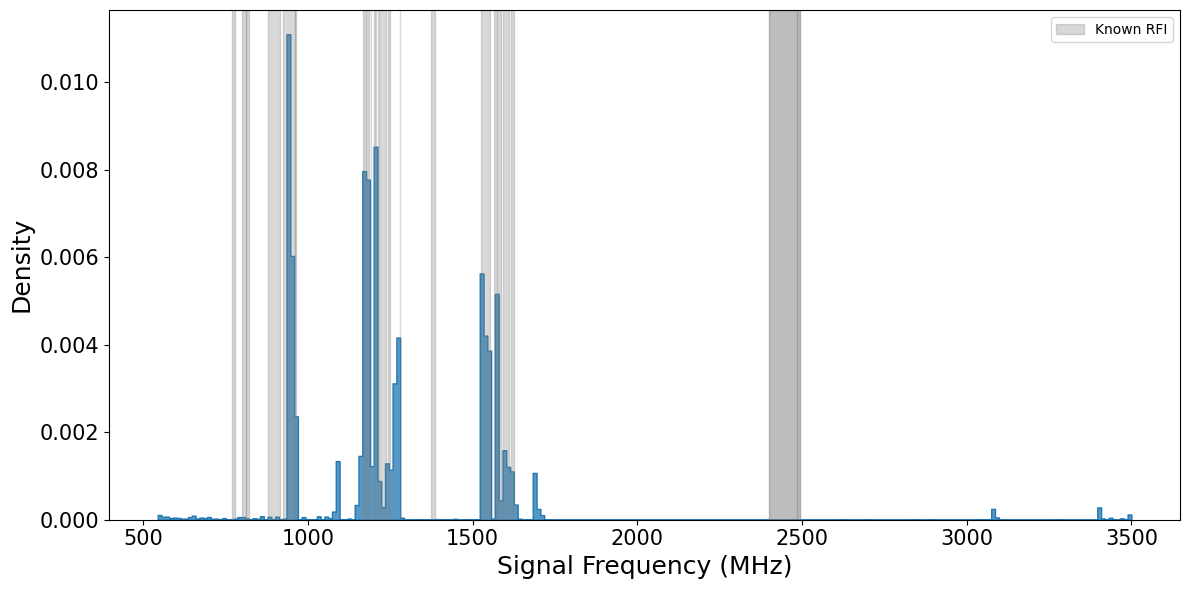}
    \caption{Histogram density plots of all the signals detected through the autonomous real-time pipeline running on BLUSE prior to RFI rejection. The plots show the data from all bands and from all of the epochs of observations. The grey regions show the RFI outlined in the MeerKAT online documentation.} 
    \label{fig:bluse-RFI}
\end{figure*}

For this analysis, we generated a separate \texttt{.csv} file for each frequency band and each telescope, specifying the start and end frequencies of all identified RFI sources and, when available, the identifier for the RFI. We then excluded all detections overlapping with these frequency intervals, independent of pointing direction. Figure~\ref{fig:c-band-RFI} and Figure~\ref{fig:bluse-RFI} illustrates this process, showing the known RFI regions for each band and telescope along with the detection of the all the signals from COSMIC and BLUSE.

An exception to this procedure was made for the \textit{S}-band (2–4,GHz) observations with the VLA. For these data, we used the \texttt{.csv} file produced in \citet{Tremblay_VLASS}, which was generated from time-averaged observations of a calibrator using an identical instrumental setup. This file incorporates both the persistent RFI detected by \textsc{COSMIC} and the supplementary information available through the \textit{NRAO} RFI database.

\subsection{Drift Rates}
\label{subsec:drift_rates}
If radio emissions from K2-18b were to be detected, the planetary signal would exhibit a Doppler drift resulting from the relative acceleration between the transmitter and the receiver. This drift manifests as a change in the observed frequency over time, caused by the combination of the planet’s orbital motion around its host star and the rotational velocity of the Earth. Following the methods described by \cite{Li_2022} and \citet{Li_2023}, we estimate that 99\% of such signals would experience a frequency drift of approximately 0.4~\,Hz\,~s$^{-1}$ for signals between 544 and 1500\,MHz, increasing to 1.879\,Hz\,s$^{-1}$ near 4.5\,GHz, and reaching a maximum of 4.177\,Hz\,s$^{-1}$ around 10\,GHz. These limits were adopted as search boundaries in our analysis.

As an additional filtering step, all signals with a drift rate of exactly 0\,Hz\,s$^{-1}$ were removed, as such detections likely correspond to local RFI, where the transmitter and receiver are either stationary relative to one another or moving too slowly to produce a measurable Doppler shift within our drift-rate resolution. 

\begin{figure}
    \centering
    \includegraphics[width=0.9\linewidth]{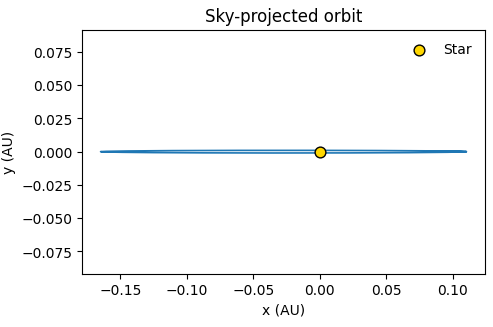}
    \caption{A Plot showing an example model of the orbital motion of K2-18b using the parameters listed in the Exoplanet Archive. This shows that the orbit is highly eccentric ($e$ = 0.2) and the transit duration is 2.66\,hours. The periastron is --5 degrees so the projection introduces a nonzero mean velocity (a constant offset).  %As the planet’s motion is not perfectly symmetric due to its eccentricity, its speed is higher at periastron than apastron, as shown in the right-hand plot. The curved line is the actual Keplerian orbital radial velocity variation. The flat line (offset) is the systemic mean velocity shift due to a term in the orbital dynamics parameter. As the longitude of ascending node is not listed within the Exoplanet archive, we set this value to zero.
    }
    \label{fig:Motion}
\end{figure}

\subsection{Strong and Weak signal excision}
We investigated the expected false-positive rate by injecting synthetic signals into recorded data using the \textsc{setigen}\footnote{\url{https://setigen.readthedocs.io/en/main/}}
 package and processing the resulting datasets with the \textsc{seticore} software, employing the same drift-rate and SNR parameters as used in the real-time search (Steigler, Tremblay, Myburgh et al., in prep.) which highlights a similar trend as discussed in \cite{Choza_2023}. The results indicate that, on average, nearly 80\% of signals with an SNR of 8\,$\sigma$ and fewer than 64 time samples were false positives—cases in which the pipeline reported a detection despite the absence of a signal in the dynamic spectrum. This was dominated by random bright pixels aligning and the de-drifted power spectrum showing a peak. Consequently, we adopted a lower SNR threshold of 10\,$\sigma$ for the \textsc{COSMIC} data, consistent with the native detection limit used in the \textsc{BLUSE} pipeline.

At the opposite extreme, any Earth-like civilization’s radio technosignature signal traversing the $\sim$124\,ly between Earth and K2-18b would potentially be weak. However, powerful transmitters could exist, a more overarching consideration are antenna-generated RFI or artifacts. To assist in determining an upper limit we analyzed 1,000 \textsc{stamp} at random containing a mix of signals from the \textsc{COSMIC} pipeline, which revealed that over 90\% of high-SNR detections were caused by instrumental effects—appearing only in a single antenna but with a high SNR—and were not consistent with sky-dependent sources. The real-time pipeline identifies these signals in the beamformed data as the signal can reach up to 3000$\sigma$ meaning that it is present in the computed data products. Therefore, we imposed a SNR upper limit of 100 during this post processing step but note that the original data detected by the pipeline is still present when a better algorithm for this type of artifact is developed. %More details will be provided in an upcoming publication.

In summary, we applied an SNR range of 10\,$\sigma$--100\,$\sigma$ to exclude both low-significance detections and spurious high-SNR events arising from strong RFI or instrumental artifacts. This will produce a limitation of detecting signals from K2-18b no brighter than 10$^{14}$\,W for COSMIC and 10$^{13}$ for BLUSE.

\subsection{Multibeam Analysis}
As shown in the right panel of Figure~\ref{fig:COSMICdata} and Figure~\ref{fig:MeerKATdata}, we simultaneously formed multiple coherent beams toward nearby \textit{Gaia} stars within the telescope’s FOV. This configuration allows us to determine whether a detected signal is spatially localized to K2-18b or instead represents RFI, which would appear in all beams at the same frequency and drift rate.

Because fine channelization is performed using a FFT, some detected signals may span multiple frequency bins if they are broader than a single channel width. To account for this, we apply a tolerance of $\pm$\,1 fine channel when cross-matching detections. For \textsc{COSMIC}, which operates with a 2\,Hz frequency resolution, this corresponds to a $\pm$\,2\,Hz tolerance. For MeerKAT, the search was completed at a 1\,Hz resolution and a 1\,Hz tolerance.

We apply a similar tolerance to the drift rate to account for standard errors in the pipeline. The \textsc{seticore} algorithm searches along all possible drift trajectories to identify the drift rate that maximizes the signal intensity, as described by \citet{Choza_2023}. For broad or nonlinearly varying signals, the derived drift rate can differ by approximately one drift step. Consequently, we adopt a tolerance of one drift step when evaluating whether detections correspond to the same signal across multiple sky positions.

\subsection{Primary and Secondary Transits}
K2-18b is a sub-Neptune-sized exoplanet located approximately 38\,pc away, residing within the habitable zone of its host star and orbiting with a period of 33\,days. We used the \textit{NASA Exoplanet Archive Transit Viewer}\footnote{\url{https://exoplanetarchive.ipac.caltech.edu/cgi-bin/TransitView}}
 to determine the epochs of both primary and secondary transits. (See Figure \ref{fig:Motion} to see a model of the orbit of the planet around the star.) Based on the model from \citet{K218b_transit}, secondary transits were predicted to occur on 2023 September 25 at 00:35, October 27 at 23:09, November 29 at 21:43, and 2024 January 1 at 20:16 (UTC). Primary transits were predicted on 2023 September 7, October 11, November 13 at 10:26, and December 16 at 08:15 (UTC). After secondary transit, the planet is occulted (Superior Conjunction) by its host star for approximately 2.7\,hours, meaning that a transmitter located on or near the planet would be visible to the telescope for 99.6\% of its orbital period.

When applicable, we compared signals detected while the planet was behind the star (secondary transit) to those observed when the planet was in front of the star. Any signal detected during the secondary transit, and immediately after while the planet is obscured from view,  cannot originate from a transmitter on or near the planet’s surface. Although it is theoretically possible for an unrelated transmitter to exist elsewhere within the system and remain out of phase with the planet, that scenario lies beyond the scope of this analysis. Using the tolerances described in Section~3.4, we identify signals exhibiting matching frequencies and drift rates while the planet is obscured by the star and when it can be observed from Earth. We retain only those detections that were unique to the planet’s visible phase and excluded all others from the candidate list.

We note that for these observations, data were not taken while K2-18b was occulted by its star. Therfore, this step is not applicable for these observations. However, as this paper also provides a framework for future observations, we have left this step as an additional filter for signals of interest. 

\subsection{Multiple Day Comparison}
When multiple observations are obtained on different days while the planet is in front of its host star, we also compare the detected signals across epochs. As described by \citet{Li_2022}, signals transmitted from a moving source (e.g., a planet’s surface) and received by a moving observer (e.g., a ground-based telescope) will experience a changing observed frequency and drift rate over time. For sufficiently long observations, the frequency evolution of such a drifting signal is expected to trace a sinusoidal pattern. Consequently, if the same source is observed on multiple days, we do not expect the signal to appear at an identical frequency or drift rate. Any detection exhibiting the same frequency and drift rate across different epochs is therefore most likely due to RFI or instrumental systematics.

Conversely, a signal that is unique in both frequency and drift rate represents the strongest technosignature candidate.

\subsection{Coherent \& Incoherent Beam Comparison}
The data generated by forming a coherent beam is targeted toward a specific source. However, an incoherent beam is a sum across the entire telescope's FOV, removing the ability to localize a signal. A signal detected in both a coherent and incoherent beam is possible, however, a signal of the same origin in the incoherent beam will have a reduced sensitivity determined by the calculation:

\begin{equation}
    \mathrm{SNR_{\mathrm{coherent}}} \leq \sqrt{N_{\mathrm{antennas}}} \times \mathrm{SNR_{\mathrm{incoherent}}}
\end{equation}

However, signals that are not of the same origin, although having a similar frequency and drift rate, will not follow this relationship. Therefore, this is a strong delineator of RFI within the field and emission emanating from a target source.

\subsection{Visual Inspection}
For our strongest technosignature candidates, we plot the dynamic spectrum and visually inspect the signals. Candidate technosignatures will be those that have passed through all of the above filters, appear in all online antennas in the array, and show a distinct linear signal drifting in time and frequency.

\section{Results}
For each telescope and receiver band the data were processed sequentially through the steps outlined in Section~\ref{PPM}. A summary of the results are found in Table \ref{tab:summary}. The observations on MeerKAT did not include an incoherent beam or multi-epoch analysis. Therefore, those steps are marked as ``Not Applicable (N/A)". For observations with both the VLA and MeerKAT the observation dates did not include when the planet was obscured by the star, so therefore that row in the table is also marked as ``N/A". Below we provide a more detailed analysis of each set of observations.

\begin{table*}
\centering
\caption{Summary Table of results from filtering the data from both COSMIC and BLUSE. The numbers are the remaining signals after each step explained in Section 3 are completed. The row ``Toward K2-18b" represents the number of signals unique to that sky position after the first three steps of filtering were completed. We use ``N/A" to denote any steps that are considered by not applicable for this set of data.}

\begin{tabular}{l|ccc|ccc}
\hline
 & \multicolumn{3}{c|}{\textbf{VLA}} & \multicolumn{3}{c}{\textbf{MeerKAT}} \\
\textbf{Step} & S band & C band & X band & UHF & L band & S4 band \\
              & 2--4\,GHz & 4--8\,GHz & 8--12\,GHz & 0.544--1.015\,GHz & 0.9--1.670\,GHz & 2.625--3.5\,GHz\\
\hline
initial          & 10,743,382 & 9,920,694 & 310,961   & 59,030 & 398,867 & 4,712 \\
Step 3.1: RFI    & 6,737,277  & 237,333   & 178,944   & 59,030 & 118,677 & 4,712 \\
Step 3.2: Drift Rates & 589,011   & 31,517    & 18,974    & 5,902  & 58,734  & 1,488 \\
Step 3.3: S \& W signals & 384,955   & 19,816    & 11,067    & 2,991  & 13,386  & 1,103 \\
Toward K2-18b    & 141,450    & 4,911     & 242       & 46     & 225     & 180 \\
Step 3.4: Multibeam     & 13,847     & 768       & 0         & 0      & 0       & 0 \\
Step 3.5: Transits      & N/A        & N/A       & N/A       & N/A    & N/A     & N/A \\
Step 3.6: Day           & 4,316      & 86        & --        & N/A    & N/A     & N/A \\
Step 3.7: Beam          & 0          & 0         & --        & N/A    & N/A     & N/A \\
Step 3.8: Visual        & 0          & 0         & --       & --    & --     & -- \\
\hline
\end{tabular}
\label{tab:summary}
\\
{\footnotesize \textit{Note.} The frequency bands listed represent the nominal receiver coverage. The exact sub-bands observed are described in the relevant sections of the text.}
\end{table*}

\subsection{COSMIC S-band Search}
All detected signal hits across the seven observing epochs were loaded, each containing frequency, SNR, drift rate, and time metadata. Known RFI frequency ranges were imported from a \texttt{.csv} file, and a mask was applied to exclude contaminated bands. Hits with $\mathrm{SNR} > 100$ were removed to avoid instrumental artifacts, reducing the dataset to $\sim$385,000 detections and resulting in only 3.5\% of the original detections remaining for analysis (Figure~\ref{fig:c-band-RFI}). This subset contained coherent, nonzero-drift signals within plausible Doppler ranges and moderate SNR values---characteristics consistent with expected astronomical rather than terrestrial origins. Unlike the other observation completed with the VLA, we observed K2-18b in S-band in seven different epochs on 03-10-2023, 05-10-2023, 13-10-2023, 22-10-2023, 8-12-2023, 14-12-2023, and 21-12-2023. In all instances the source was visible from Earth.

The filtered data were divided into three groups: signals detected toward K2-18b, those from other field sources detected in coherent beams, and signals detected in the incoherent sum. {Candidates that are unique to the coherent beam toward K2-18b were identified by requiring frequency differences greater than $\pm 2$\,Hz and drift-rate differences exceeding $\pm 0.1$\,Hz\,s$^{-1}$ relative to all other detections, resulting in 13,847 unique signals.

Finally, these detections were classified by date using the \texttt{tstart} timestamp. After analysis 663 candidates matched between the incoherent and coherent beams, however, none showed an expected power ratio (4.69) for a real astrophysical signal. After applying all remaining filters described in Section~\ref{PPM}, no viable technosignature candidates were identified.

\subsection{COSMIC C-band Search}
The COSMIC system recorded observations of K2-18b over three consecutive weeks on 2023 December 8, 14, and 19 at frequencies between 5 and 7.2\,GHz, resulting in 9,920,694 detected hits. Each observation consisted of 10 minutes on-source, divided into 56\,s segments that were searched independently. Data for December 8 and 14 were obtained after K2-18b was out of secondary transit, whereas observations for December 19 occurred shortly after primary transit on December 16. However, at no time during these observations was the planet obscured from view. A histogram of the hits detected by the real-time pipeline is shown in Figure~\ref{fig:c-band-RFI}.

In the first stage of processing, the known RFI frequency bins were removed using a predefined mask and the remaining signals were filtered as described in Section~\ref{PPM}. After applying drift-rate limits of $\pm$1.9\,Hz\,s$^{-1}$ and restricting detections to those with $10 \leq \mathrm{SNR} \leq 100$, less than 5,000 candidate signals towards K2-18b remained. Of those signals, no signals had a match between the incoherent beam and the coherent beam from step 3.7, so all remaining signals from the coherent beam were searched for a comparison of signals between each of the three days of observations (Step 3.6). Following the post-processing pipeline, 86 signals were retained for visual inspection using the \textit{stamp} files. The dynamic spectra for each remaining signal had similar drift rate and angles. Of these remaining signals, 90\% of the signals were within a MHz of each other around 6.6\,GHz. The spectra for each antenna showed inconsistency (not present in all antennas and not present in all time bins), suggesting that all remaining signals were RFI instead of a signal originating from the K2-18 system. The dynamic spectrum of four examples of the remaining signals, all found to be RFI or telescope artifacts, are shown in Figure \ref{fig:Cbandsignals}.

\begin{figure*}
    \centering

    % Top row
    \begin{minipage}[t]{0.48\linewidth}
        \centering
        \begin{overpic}[width=\linewidth]{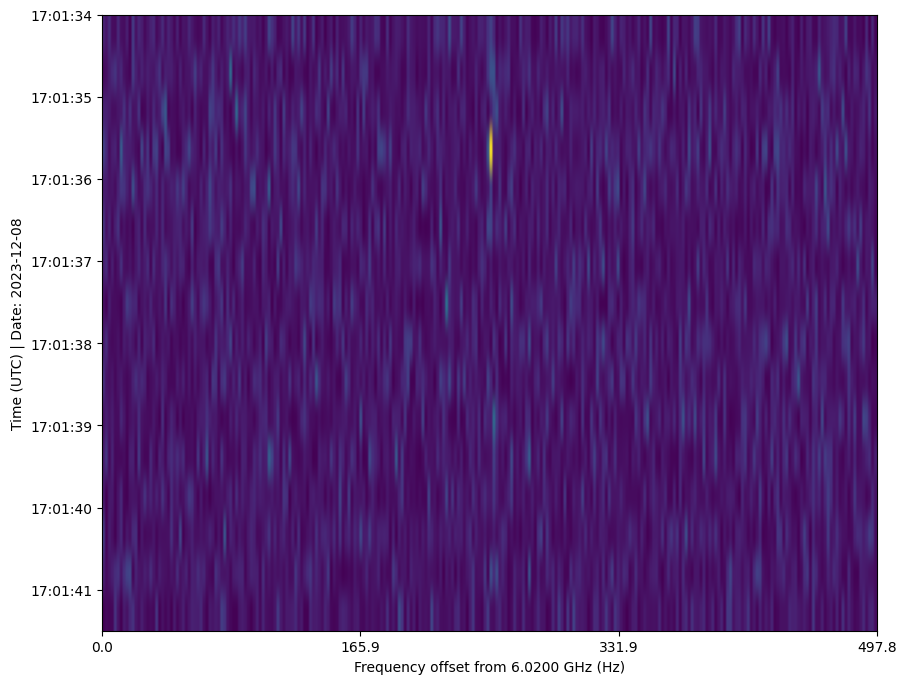}
            \put(48,-5){\textbf{(a)}} % near top-right
        \end{overpic}
    \end{minipage}
    \hfill
    \begin{minipage}[t]{0.48\linewidth}
        \centering
        \begin{overpic}[width=\linewidth]{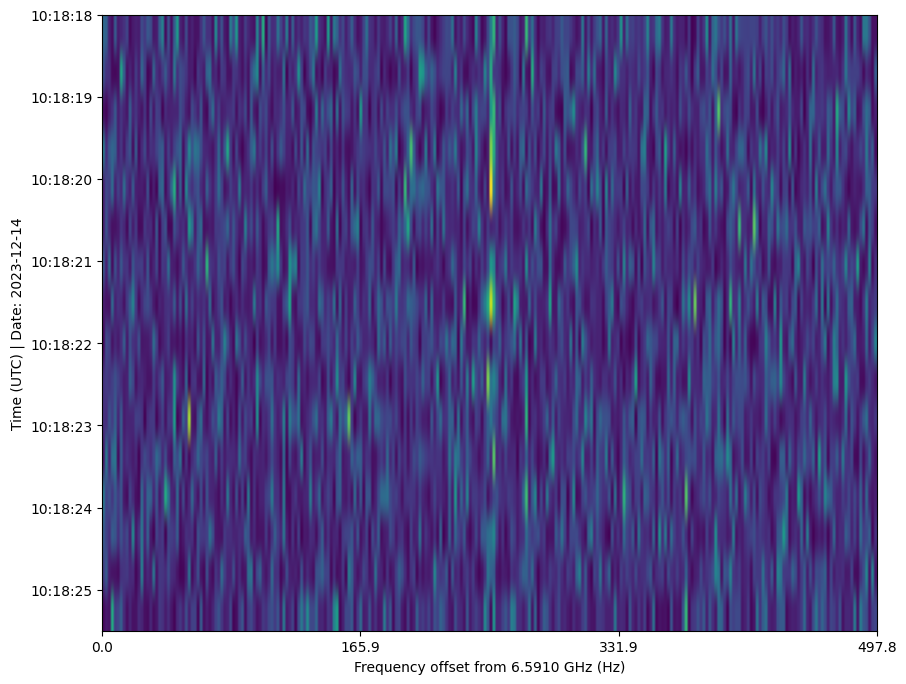}
            \put(48,-5){\textbf{(b)}}
        \end{overpic}
    \end{minipage}

    \vspace{0.3cm}

    % Bottom row
    \begin{minipage}[t]{0.48\linewidth}
        \centering
        \begin{overpic}[width=\linewidth]{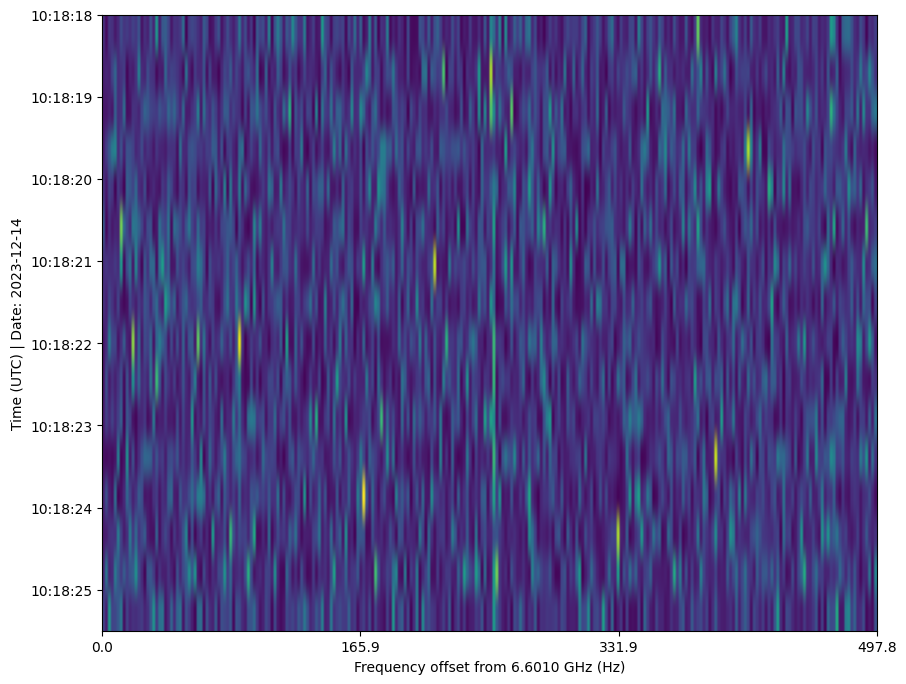}
            \put(48,-5){\textbf{(c)}}
        \end{overpic}
    \end{minipage}
    \hfill
    \begin{minipage}[t]{0.48\linewidth}
        \centering
        \begin{overpic}[width=\linewidth]{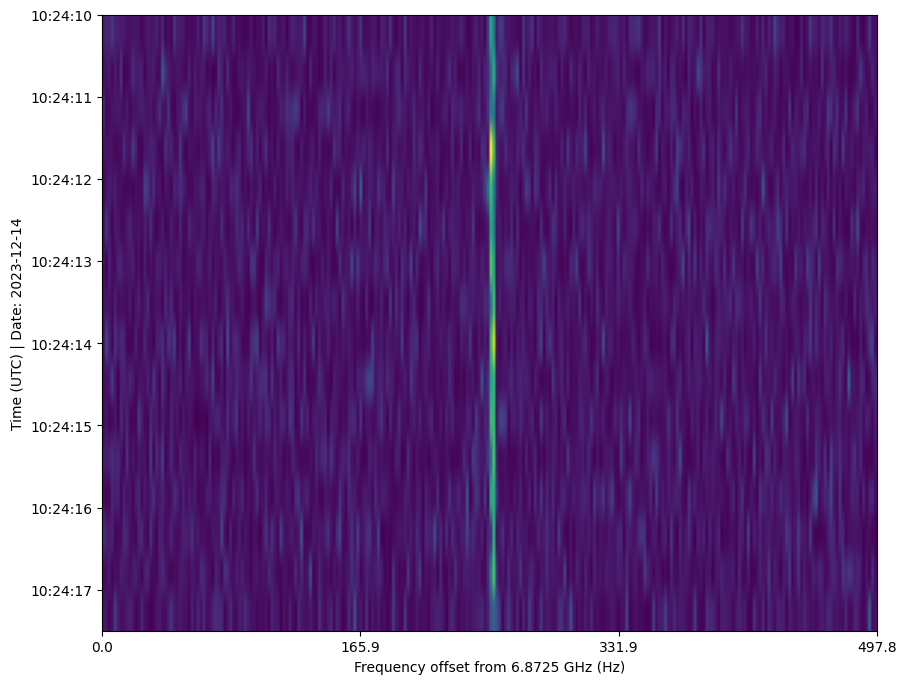}
            \put(48,-5){\textbf{(d)}}
        \end{overpic}
    \end{minipage}

    \caption{Dynamic spectrum of four signals visualized after they passed through the criteria listed in the pipeline. All signals were found to be RFI or an instrument artifact. (a) Spectra showing a bright pixel with a weak drifting signal. The signal was not present in all antennas. (b) A weak signal drifting in time and frequency was an inconsistent intensity. The signal was not present in all antennas. (c) An example weak signal in a noisy part of the spectrum and surrounded by RFI.  (d) A bright drifting signal only present in one antenna, representing a instrument artifact.}
    \label{fig:Cbandsignals}
\end{figure*}

\subsection{COSMIC X-band Search}
Similarly to the C-band observations, the COSMIC system recorded additional data over three consecutive weeks on 2023 December 8, 14, and 19. These observations were conducted across the same four coherent beams, each targeting different sources within the FOV, including K2-18b. In total, the real-time pipeline detected 310,961 hits over these three epochs. Each observation consisted of 10 minutes on-source, divided into 56\,s segments that were independently calibrated, beamformed, and searched for narrowband drifting signals.

Following the post-processing workflow described in Section~\ref{PPM}, all detections were first cross-matched against known RFI frequency bins, and hits located within contaminated bands were removed. The remaining signals were filtered based on drift rate, retaining only those within $\pm$4.2\,Hz\,s$^{-1}$ and with SNR between 10 and 100. After applying these constraints, 242 hits remained that were unique to the position of K2-18b.

These 242 candidate detections were then examined for spatial consistency across the four coherent beams by comparing their appearance and intensity. A genuine technosignature would be expected to appear only in the beam corresponding to the K2-18b position and not in the off-target data. However, after this comparison, no surviving signals exhibited spatial isolation or consistent drift-rate behavior, indicative of an astrophysical or artificial origin. As a result, no technosignature candidates were identified in the observations.

\subsection{MeerKAT UHF}
Because the BLUSE observations in the UHF band were conducted on a single day, we relied primarily on the spatial filtering afforded by the backend's 64 coherent beams to distinguish RFI from potential technosignatures. Each beam targeted a different position on the sky drawn from a catalog of nearby stellar systems, including K2-18b. This configuration allows local interference to be recognized through its appearance in multiple beams, whereas an astrophysical signal would be confined to only one.

Across the observing period, the real-time detection system reported 59,030 hits within the 544--1015\,MHz band. Drift rates ranged from $-10$ to $+9$\,Hz\,s$^{-1}$, with $\approx 83\%$ of detections exhibiting zero drift, consistent with stationary terrestrial emitters. After applying the masking procedure described in Section~\ref{PPM} and discarding signals outside the allowed drift-rate window, we further restricted the dataset to detections with $\mathrm{SNR}$ between 10 and 100, leaving 5.1\% of the initial hit set.

A final spatial comparison required that any surviving detection appear exclusively in the K2-18b beam and in none of the other 63 beams. No signals met this criterion and no viable technosignature candidates were identified in the UHF data set.

\subsection{MeerKAT L-band}
For the L-band observations, the BLUSE system again synthesized 64 coherent beams across the MeerKAT field, each directed toward a unique entry in the predefined target catalog. This multi-beam strategy provides strong spatial discrimination against RFI, as genuine extraterrestrial transmitters are expected to appear in only a single beam.

The BLUSE pipeline identified 398,867 initial detections during the L-band session. Masking frequencies associated with persistent RFI reduced the sample to 118,677 detections. Imposing the standard drift-rate constraint of $\pm 1.9$\,Hz\,s$^{-1}$ further reduced the set to 58{,}734 signals, and limiting the selection to $10 \leq \mathrm{SNR} \leq 100$ yielded 13,386 hits appropriate for spatial analysis.

Prior to cross-matching these detections across all 64 coherent beams, 225 signals appeared solely in the K2-18b beam. However, no signals toward K2-18b were unique when compared to the other 63 coherent beams, so no further analysis was required.

\subsection{MeerKAT S-band}

The real-time pipeline detected a total of 4,712 hits across the S4 observing band. None of these overlapped with frequencies flagged in the catalog of persistent RFI, and thus no detections were removed at this preliminary stage.

Applying the nominal drift-rate constraint of $\pm 1.9$\,Hz\,s$^{-1}$ reduced the dataset to 1,488 signals, and further restricting the sample to detections with $10 \leq \mathrm{SNR} \leq 100$ yielded 1{,}103 hits for spatial comparison. Each of these was examined for exclusivity to the beam centered on K2-18b.

No detections were found to be unique to the K2-18b beam; all surviving signals were either coincident across multiple beams or exhibited characteristics indicative of RFI. As a result, no technosignature candidates were identified in the S-band observations.

\section{Discussion}
\label{sec:discussion}

\subsection{Constraints on Technosignatures in the K2--18 System}

We have conducted a coordinated, multi-epoch search for narrowband technosignatures toward the K2--18 system using two interferometric facilities: the VLA with the \textsc{COSMIC} backend and the MeerKAT array with the \textsc{BLUSE} backend. Complementary analyses of our VLA observations show no coherent or incoherent stellar radio emission from K2-18 and place $\mu$Jy-level upper limits on its magnetic activity \citep{wandia2025upper}. After applying a sequence of filters based on RFI excision, drift-rate constraints, multibeam spatial localization, primary transit behavior, and SNR thresholds, no viable technosignature candidates were identified in any band or epoch.

This null result allows us to place upper limits on continuously emitting, narrowband transmitters within the K2--18 system. At a distance $\mathrm{d_p}$=38\,pc\footnote{\url{https://exoplanetarchive.ipac.caltech.edu/overview/K2-18}}, the sensitivities achieved in our observations imply that any isotropic transmitter located on or near K2--18b and emitting between 544\,MHz and 9.8\,GHz must have an EIRP below the limits derived in this work (see Section~\ref{PPM} and Table~\ref{tab:eirp_limits}). All of these values are close to or less than the power provided by the Arecibo telescope ($\sim 10^{13}$\,W; \citealt{Siemion_2013}) and put stringent constraints on the system. These constraints apply to transmitters that are narrowband (a few Hz wide), persistent on timescales comparable to our integration times, and emitting such that the Earth lies within the beam pattern during the epochs observed. Intermittent, broadband, or highly directional signals that did not illuminate the Earth at those times are not ruled out by this study.

Although we do not detect technosignatures, these limits contribute to the growing set of observational constraints on technologically active civilizations around nearby stars and represent one of the first targeted searches toward a Hycean-planet candidate.

\begin{deluxetable}{lccc}
\tablecaption{EIRP Limits for K2-18b for each receiver band observation. \label{tab:eirp}}
\tablehead{
\colhead{Telescope} &
\colhead{Band} &
\colhead{Frequency Range (MHz)} &
\colhead{EIRP Limit (10$^{12}$ W)}
}
\startdata
%--------------------------------------------------
% Fill in your values here
% Example format:
% MeerKAT & UHF & 544--1015 & $X$ \\
% VLA     & S   & 2400--3600 & $Y$ \\
%--------------------------------------------------
MeerKAT & UHF & 544--1015 &  1.6\\
 MeerKAT & L & 900--1670 & 1.2 \\
 MeerKAT& S4 & 2625--3500  & 1.1 \\
VLA & S & 2500--3500 & 23 \\
VLA & C & 5800--7200 &  16\\
VLA & X & 8000--9800 &  13\\
\enddata
\tablecomments{EIRP limits correspond to the minimum detectable isotropic transmitter power at the distance of K2-18b, using the sensitivity of each observing band and assuming narrowband emission. See the Appendix for details on the calculation.}
\label{tab:eirp_limits}
\end{deluxetable}

\subsection{Implications for K2--18b and Hycean Worlds}

K2--18b is an important benchmark for studies of sub-Neptune atmospheres and Hycean-world scenarios. Atmospheric inferences from JWST suggest a hydrogen-rich atmosphere with CH$_4$ and CO$_2$ and a lack of NH$_3$ and CO, with important implications for a water-rich interior and a possible Hycean world \citep{Madhusudhan_2023, Hu_2025, Howard_2025}. As outlined in detail by \cite{Wright_2022}, our radio observations provide an independent constraint. Within the narrowband, low-drift parameter space probed here, we find no evidence for technologically generated radio emission from the K2--18 system.

This result is consistent with several possibilities: K2--18b may be uninhabited, it may host pre-technological life, or any technologically capable civilization may employ communication modalities that are not detectable with our current observing strategy (e.g., non-radio, broadband, low-duty cycle, or highly encrypted signals). Thus, the absence of a detection should not be interpreted as evidence against habitability or against the presence of technological activity in any absolute sense. Rather, it constrains a specific class of transmitters: persistent, narrowband emitters in the observed frequency range.

Nevertheless, combining atmospheric characterization with direct technosignature limits is valuable. K2--18b illustrates how detailed exoplanet atmospheric studies and radio technosignature searches can be coordinated, particularly for systems in or near the habitable zone. As more Hycean candidates are identified, similar joint strategies will help place such systems in a broader comparative context.

\subsection{Methodological Considerations and the Role of Interferometers}

A key outcome of this work is the demonstration of a general post-processing framework for interferometric technosignature searches. Our approach integrates: (1) real-time calibration, channelization, beamforming, and drift-rate searches via \textsc{COSMIC} and \textsc{BLUSE}; (2) observatory-informed and empirically derived RFI masks; (3) multibeam spatial filtering to distinguish localized sky signals from field-wide interference; (4) drift-rate filtering based on the expected dynamics of the K2--18 system; and (5) orbital-phase-based filtering using primary and secondary transits.

Although this is not the first study toward exoplanets \citep{Sheikh_2023_keppler,Barrett_2025,Saide_2023}, the multibeam capability used in this paper is particularly important. A genuine signal from the K2--18 system is expected to appear in a single coherent beam, with an amplitude consistent with the incoherent sum (Section 3.7), whereas local RFI often appears in multiple beams and does not satisfy the expected coherent-to-incoherent SNR relation. This spatial discrimination, combined with transit-based and multi-epoch consistency checks, substantially reduces the false-positive rate that was built off from the framework published in \cite{Tusay_2024} but modified to account for the larger number of coherent beams distributed across the telescope FOV.

These techniques are not specific to K2--18b. The framework can be applied to other exoplanetary systems observed with commensal interferometric backends and is readily scalable to larger target samples. As widefield interferometers and digital backends become more prevalent, such multibeam, multi-epoch strategies will play an increasingly central role in radio technosignature searches.

\subsection{Limitations and Future Prospects}

Several limitations of this study should be noted. First, although our observations span at least one orbital period, they sample only a small fraction of the planet’s possible orbital phases and any potential transmission strategies. Low-duty-cycle or phase-specific transmitters could be missed. Second, the search parameter space is restricted to approximately linear drift rates within the ranges motivated by the expected accelerations of the K2--18 system; signals with strongly nonlinear or rapidly varying frequency evolution may not be efficiently recovered. Third, while our RFI models combine observatory information and empirical masking, they are necessarily incomplete. Some low-level or intermittent interference may remain, and conservative masking can exclude regions in which genuine signals might reside.

Future work can address these limitations in several ways. Additional observations of K2--18b with improved temporal sampling and broader frequency coverage would increase sensitivity to intermittent or phase-dependent transmissions. Incorporating polarization information, more flexible drift models, and machine-learning-based anomaly detection into the pipeline may enhance sensitivity to nonstandard signal morphologies. Finally, applying the methods developed here to a larger sample of targets—particularly other nearby Hycean candidates and habitable-zone planets—will help place the K2--18 limits in a statistical context.

\subsection{Broader Context}

Technosignature searches remain exploratory, but carefully designed null results provide meaningful constraints on the prevalence and characteristics of technologically active civilizations. For K2--18b, we present the first interferometric search of this system for narrowband radio emission over a wide frequency range, complementing ongoing atmospheric studies and establishing a template for future coordinated campaigns.

As next-generation facilities such as the Square Kilometer Array (SKA) and Next-generation VLA (ngVLA) come online, the strategies used here—multibeam searches, transit- and drift-aware filtering, and systematic post-processing—will be essential for scaling technosignature searches to larger datasets and higher sensitivities. K2--18b thus serves both as a scientifically compelling target and as a test case for the methodologies required in future large-scale searches for technosignatures.

\section{Conclusion}
In this work, we conducted a coordinated, multi-facility search for narrowband technosignatures in the K2–18 system using the VLA \textsc{COSMIC} and MeerKAT \textsc{BLUSE} backends, spanning 544\,MHz to 9.8\,GHz and covering multiple epochs across one orbital period of K2–18b. Using a uniform post-processing framework that incorporated observatory-informed RFI excision, drift-rate filtering based on the planet’s orbital dynamics, multibeam spatial discrimination, cross-epoch consistency checks, and primary/secondary transit constraints, we identified no signals consistent with an artificial origin. All surviving detections were attributable to terrestrial RFI or instrumental systematics.

These non-detections enable the first expansive search for interferometric technosignature limits for a proposed Hycean-world candidate. The absence of persistent narrowband emission at our achieved sensitivities places upper limits on isotropic transmitters within the K2–18 system and demonstrates the value of combining independent arrays, backend architectures, and observing strategies. %Our results highlight both the challenges and opportunities inherent in technosignature searches near magnetically active M dwarfs, where stellar variability—or planetary occultation—can provide astrophysical context absent in broad Galactic surveys.

The methodology presented here establishes a generalizable framework for future targeted searches of potentially habitable exoplanets with current and next-generation facilities such as the ngVLA and SKA. Continued multi-epoch, multi-band observations of nearby sub-Neptunes will further refine technosignature constraints, complement atmospheric studies, and expand the parameter space in which we can meaningfully search for technological activity beyond the Solar System.

\section{Data Availability}
All data were collected via commensal backends on the VLA and MeerKAT telescopes and are under the ownership of Breakthrough Listen and the SETI Institute. The voltage, filterbank, or database information can be made available upon request to the authors. It is our plan in the future to make this data publicly available through a web service.

%% Also note that the akcnowlodgment environment does not support long amounts of text. If you have a lot of people and institutions to acknowledge, do not use this command. Instead, create a new 
\section{Acknowledgments}
%\begin{acknowledgments}
The authors gratefully acknowledge the foundational support from John and Carol Giannandrea that made COSMIC possible. As part of the Breakthrough Listen project, the BLUSE system on the MeerKAT telescope is sponsored by Breakthrough Initiatives, an organization affiliated with the Breakthrough Prize Foundation. The National Radio Astronomy Observatory is a facility of the National Science Foundation operated under a cooperative agreement with Associated Universities, Inc. The MeerKAT telescope is operated by the South African Radio Astronomy Observatory, which is a facility of the National Research Foundation, an agency of the Department of Science and Innovation. This work has made use of the “MPIfR S-band receiver system” designed, constructed and maintained by funding of the MPI für Radioastronomy and the Max-Planck-Society.
%\end{acknowledgments}


\vspace{5mm}
\facilities{VLA, MeerKAT}


\software{astropy,pandas,seaborn \citep{2013A&A...558A..33A,2018AJ....156..123A,reback2020pandas,Waskom2021}} 
          
\appendix          
\section{Sensitivity Limit Calculation}
In both the COSMIC and BLUSE experiments the signals are searched for in beamformed data. Therefore, the sensitivity limits, when no signals originating from the planet are detected, are carefully considered. To set the flux density limit (S$_{limit}$) we use the standard equation for beamformed data:

\begin{equation}
S_{limit} = \frac{ \mathrm{SEFD}}{B_e\sqrt{n_{pol} \times n \times t_{int} \times \Delta \nu}} \times SNR
\end{equation}

The system equivalent flux density (SEFD) is determined using the standard radiometer equation and limits published on both the NRAO\footnote{\url{https://science.nrao.edu/facilities/vla/docs/manuals/oss/performance/sensitivity}} website for COSMIC and SARAO\footnote{\url{https://skaafrica.atlassian.net/wiki/spaces/ESDKB/pages/277315585/MeerKAT+specifications}} website for BLUSE and is shown in Equation A2. The value of $S_{limit}$ are in Jy and are converted using the definition of 1\,Jy is equal to 1$\times$10$^{-26}$\,W/m$^2$/Hz.

\begin{equation}
\mathrm{SEFD} = \frac{2 k_B T_{sys}}{A_{eff}}
\end{equation}

The beamformer efficiency value (B$_e$) is determined through observations on each system. In \cite{tremblay_cosmic}, we determined the beamformer efficiency on COSMIC using a methanol maser to be between 0.85 and 0.95 depending on the day. For BLUSE, we used observations of the signals from JWST to determine a beamformer efficiency of 0.93. Therefore, we use a value of B$_{e}$ of 0.9 for COSMIC and 0.93 for BLUSE.

The values for the number of Polarizations (n$_{pol}$), the number of antennas (n), the integration time (t$_{int}$, SNR as the threshold of the experiment, and the bandwidth ($\Delta \nu$) are from Table \ref{tab:obs_setup}. These value will potentially change per each experiment. We use the bandwidth of the single channel width, as we are looking for single constrained to a single channel.

To compare our limits to other technosignature surveys and to what these sensitivity limits mean for the detection of a transmitter, we use the equation for Equivalent Isotropic Radio Power (EIRP).

\begin{equation}
\mathrm{EIRP} = 4 \pi \mathrm{F}_{min} \times D^2 
\end{equation}

The value $D$ is the distance to the planet in meters. The value F$_{min}$ is the $S_{limit}$ times the single channel bandwidth (or spectral resolution).

%\newpage

\bibliography{cosmic}{}

@article{Waskom2021, doi = {10.21105/joss.03021}, url = {https://doi.org/10.21105/joss.03021}, year = {2021}, publisher = {The Open Journal}, volume = {6}, number = {60}, pages = {3021}, author = {Waskom, Michael L.}, title = {seaborn: statistical data visualization}, journal = {Journal of Open Source Software} }

@software{reback2020pandas,
  author = {The pandas development team},
  title = {pandas-dev/pandas: Pandas},
  month = feb,
  year = {2020},
  publisher = {Zenodo},
  version = {latest},
  doi = {10.5281/zenodo.3509134},
  url = {https://doi.org}
}

@ARTICLE{Sheikh_2025,
       author = {{Sheikh}, Sofia Z. and {Huston}, Macy J. and {Fan}, Pinchen and {Wright}, Jason T. and {Beatty}, Thomas and {Martini}, Connor and {Kopparapu}, Ravi and {Frank}, Adam},
        title = "{Earth Detecting Earth: At What Distance Could Earth's Constellation of Technosignatures Be Detected with Present-day Technology?}",
      journal = {\aj},
         year = 2025,
        month = feb,
       volume = {169},
       number = {2},
          eid = {118},
        pages = {118},
          doi = {10.3847/1538-3881/ada3c7},
archivePrefix = {arXiv},
       eprint = {2502.02614},
 primaryClass = {astro-ph.IM},
       adsurl = {https://ui.adsabs.harvard.edu/abs/2025AJ....169..118S}
}

@ARTICLE{Fan_2025,
       author = {{Fan}, Pinchen and {Wright}, Jason T. and {Lazio}, T. Joseph W.},
        title = "{Detecting Extraterrestrial Civilizations that Employ an Earth-level Deep Space Network}",
      journal = {\apjl},
         year = 2025,
        month = sep,
       volume = {990},
       number = {1},
          eid = {L1},
        pages = {L1},
          doi = {10.3847/2041-8213/adf6b0},
archivePrefix = {arXiv},
       eprint = {2508.15425},
 primaryClass = {astro-ph.IM},
       adsurl = {https://ui.adsabs.harvard.edu/abs/2025ApJ...990L...1F}
}

@ARTICLE{Tusay_2024,
       author = {{Tusay}, Nick and {Sheikh}, Sofia Z. and {Sneed}, Evan L. and {Farah}, Wael and {Pollak}, Alexander W. and {Cruz}, Luigi F. and {Siemion}, Andrew and {DeBoer}, David R. and {Wright}, Jason T.},
        title = "{A Radio Technosignature Search of TRAPPIST-1 with the Allen Telescope Array}",
      journal = {\aj},
         year = 2024,
        month = dec,
       volume = {168},
       number = {6},
          eid = {283},
        pages = {283},
          doi = {10.3847/1538-3881/ad823c},
archivePrefix = {arXiv},
       eprint = {2409.08313},
 primaryClass = {astro-ph.EP},
       adsurl = {https://ui.adsabs.harvard.edu/abs/2024AJ....168..283T}
}

@ARTICLE{cloutier2019,
       author = {{Cloutier}, R. and {Astudillo-Defru}, N. and {Doyon}, R. and
         {Bonfils}, X. and {Almenara}, J. -M. and {Bouchy}, F. and
         {Delfosse}, X. and {Forveille}, T. and {Lovis}, C. and {Mayor}, M. and
         {Menou}, K. and {Murgas}, F. and {Pepe}, F. and {Santos}, N.~C. and
         {Udry}, S. and {W{\"u}nsche}, A.},
        title = "{Confirmation of the radial velocity super-Earth K2-18c with HARPS and CARMENES}",
      journal = {Astronomy \& Astrophysics},
         year = "2019",
        month = "Jan",
       volume = {621},
          eid = {A49},
        pages = {A49},
          doi = {10.1051/0004-6361/201833995},
archivePrefix = {arXiv},
       eprint = {1810.04731},
 primaryClass = {astro-ph.EP},
       adsurl = {https://ui.adsabs.harvard.edu/abs/2019A&A...621A..49C}
}

@ARTICLE{Benneke2019,
       author = {{Benneke}, Bj{\"o}rn and {Wong}, Ian and {Piaulet}, Caroline and {Knutson}, Heather A. and {Lothringer}, Joshua and {Morley}, Caroline V. and {Crossfield}, Ian J.~M. and {Gao}, Peter and {Greene}, Thomas P. and {Dressing}, Courtney and {Dragomir}, Diana and {Howard}, Andrew W. and {McCullough}, Peter R. and {Kempton}, Eliza M. -R. and {Fortney}, Jonathan J. and {Fraine}, Jonathan},
        title = "{Water Vapor and Clouds on the Habitable-zone Sub-Neptune Exoplanet K2-18b}",
      journal = {The Astrophysical Journal Letters},
         year = 2019,
        month = dec,
       volume = {887},
       number = {1},
          eid = {L14},
        pages = {L14},
          doi = {10.3847/2041-8213/ab59dc},
archivePrefix = {arXiv},
       eprint = {1909.04642},
 primaryClass = {astro-ph.EP},
       adsurl = {https://ui.adsabs.harvard.edu/abs/2019ApJ...887L..14B}
}

@article{Madhusudhan_2020,
	doi = {10.3847/2041-8213/ab7229},
	url = {https://doi.org/10.3847%2F2041-8213%2Fab7229},
	year = 2020,
	month = {feb},
	publisher = {American Astronomical Society},
	volume = {891},
	number = {1},
	pages = {L7},
	author = {Nikku Madhusudhan and Matthew C. Nixon and Luis Welbanks and Anjali A. A. Piette and Richard A. Booth},
	title = {The Interior and Atmosphere of the Habitable-zone Exoplanet K2-18b},
	journal = {The Astrophysical Journal}}

@article{Madhusudhan_2021,
    title = {Habitability and {Biosignatures} of {Hycean} {Worlds}},
    volume = {918},
    issn = {0004-637X},
    url = {https://dx.doi.org/10.3847/1538-4357/abfd9c},
    doi = {10.3847/1538-4357/abfd9c},
    language = {en},
    number = {1},
    urldate = {2023-11-29},
    journal = {The Astrophysical Journal},
    author = {Madhusudhan, Nikku and Piette, Anjali A. A. and Constantinou, Savvas},
    month = aug,
    year = {2021},
    note = {Publisher: The American Astronomical Society},
    pages = {1},
}

@ARTICLE{Hu_2021,
       author = {{Hu}, Renyu and others},
        title = "{Unveiling Shrouded Oceans on Temperate sub-Neptunes via Transit Signatures of Solubility Equilibria versus Gas Thermochemistry}",
      journal = {Astrophys. J. Lett. },
         year = 2021,
        month = nov,
       volume = {921},
       number = {1},
          eid = {L8},
        pages = {L8},
          doi = {10.3847/2041-8213/ac1f92},
archivePrefix = {arXiv},
       eprint = {2108.04745},
 primaryClass = {astro-ph.EP},
       adsurl = {https://ui.adsabs.harvard.edu/abs/2021ApJ...921L...8H}
}

@ARTICLE{Tsai_2021,
       author = {{Tsai}, Shang-Min and others},
        title = "{Inferring Shallow Surfaces on Sub-Neptune Exoplanets with JWST}",
      journal = {Astrophys. J. Lett. },
         year = 2021,
        month = dec,
       volume = {922},
       number = {2},
          eid = {L27},
        pages = {L27},
          doi = {10.3847/2041-8213/ac399a},
archivePrefix = {arXiv},
       eprint = {2111.06429},
 primaryClass = {astro-ph.EP},
       adsurl = {https://ui.adsabs.harvard.edu/abs/2021ApJ...922L..27T}
}

@misc{Welbanks_2025,
    title = {The {Challenges} of {Detecting} {Gases} in {Exoplanet} {Atmospheres}},
    url = {http://arxiv.org/abs/2504.21788},
    doi = {10.48550/arXiv.2504.21788},
    urldate = {2025-05-01},
    publisher = {arXiv},
    author = {Welbanks, Luis and Nixon, Matthew C. and McGill, Peter and Tilke, Lana J. and Wiser, Lindsey S. and Rotman, Yoav and Mukherjee, Sagnick and Feinstein, Adina and Line, Michael R. and Seager, Sara and Beatty, Thomas G. and Seligman, Darryl Z. and Parmentier, Vivien and Sing, David},
    month = apr,
    year = {2025},
    note = {arXiv:2504.21788 [astro-ph]},
}

@article{Madhusudhan_2025,
    title = {New {Constraints} on {DMS} and {DMDS} in the {Atmosphere} of {K2}-18 b from {JWST} {MIRI}},
    volume = {983},
    issn = {2041-8205},
    url = {https://dx.doi.org/10.3847/2041-8213/adc1c8},
    doi = {10.3847/2041-8213/adc1c8},
    language = {en},
    number = {2},
    urldate = {2025-04-24},
    journal = {The Astrophysical Journal Letters},
    author = {Madhusudhan, Nikku and Constantinou, Savvas and Holmberg, Måns and Sarkar, Subhajit and Piette, Anjali A. A. and Moses, Julianne I.},
    month = apr,
    year = {2025},
    note = {Publisher: The American Astronomical Society},
    pages = {L40},
}

@ARTICLE{Pica-Ciamarra_2025,
       author = {{Pica-Ciamarra}, Lorenzo and {Madhusudhan}, Nikku and {Cooke}, Gregory J. and {Constantinou}, Savvas and {Binet}, Martin},
        title = "{A Systematic Search for Trace Molecules in Exoplanet K2-18 b}",
      journal = {arXiv e-prints},
         year = 2025,
        month = may,
          eid = {arXiv:2505.10539},
        pages = {arXiv:2505.10539},
          doi = {10.48550/arXiv.2505.10539},
archivePrefix = {arXiv},
       eprint = {2505.10539},
 primaryClass = {astro-ph.EP},
       adsurl = {https://ui.adsabs.harvard.edu/abs/2025arXiv250510539P}
}

@ARTICLE{Barrier_2025,
       author = {{Barrier}, Edouard F.~L. and {Madhusudhan}, Nikku},
        title = "{General circulation models of Hycean worlds}",
      journal = {\mnras},
         year = 2025,
        month = dec,
       volume = {544},
       number = {4},
        pages = {4098-4118},
          doi = {10.1093/mnras/staf1948},
archivePrefix = {arXiv},
       eprint = {2511.07546},
 primaryClass = {astro-ph.EP},
       adsurl = {https://ui.adsabs.harvard.edu/abs/2025MNRAS.544.4098B}
}

@ARTICLE{Hu_2025,
       author = {{Hu}, Renyu and {Bello-Arufe}, Aaron and {Tokadjian}, Armen and {Yang}, Jeehyun and {Damiano}, Mario and {Roy}, Pierre-Alexis and {Coulombe}, Louis-Philippe and {Madhusudhan}, Nikku and {Constantinou}, Savvas and {Benneke}, Bj{\"o}rn},
        title = "{A water-rich interior in the temperate sub-Neptune K2-18 b revealed by JWST}",
      journal = {arXiv e-prints},
         year = 2025,
        month = jul,
          eid = {arXiv:2507.12622},
        pages = {arXiv:2507.12622},
          doi = {10.48550/arXiv.2507.12622},
archivePrefix = {arXiv},
       eprint = {2507.12622},
 primaryClass = {astro-ph.EP},
       adsurl = {https://ui.adsabs.harvard.edu/abs/2025arXiv250712622H}
}

@ARTICLE{Sheikh_2023_keppler,
       author = {{Sheikh}, Sofia Z. and {Kanodia}, Shubham and {Lubar}, Emily and {Bowman}, William P. and {Ca{\~n}as}, Caleb I. and {Gilbertson}, Christian and {MacDonald}, Mariah G. and {Wright}, Jason and {MacMahon}, David and {Croft}, Steve and {Price}, Danny and {Siemion}, Andrew and {Drew}, Jamie and {Worden}, S. Pete and {Trenholm}, Elizabeth and {The Breakthrough Listen Initiative}},
        title = "{A Green Bank Telescope Search for Narrowband Technosignatures between 1.1 and 1.9 GHz During 12 Kepler Planetary Transits}",
      journal = {\aj},
         year = 2023,
        month = feb,
       volume = {165},
       number = {2},
          eid = {61},
        pages = {61},
          doi = {10.3847/1538-3881/aca907},
archivePrefix = {arXiv},
       eprint = {2212.05137},
 primaryClass = {astro-ph.EP},
       adsurl = {https://ui.adsabs.harvard.edu/abs/2023AJ....165...61S}
}

@ARTICLE{Barrett_2025,
       author = {{Barrett}, Rebecca A.~W. and {Tremblay}, Chenoa D. and {Price}, Danny C. and {Green}, Jimi A. and {Addison}, Brett C.},
        title = "{Breakthrough listen: A technosignature search around 27 eclipsing exoplanets selected from the Transiting Exoplanet Survey Satellite catalogue}",
      journal = {\pasa},
         year = 2025,
        month = jun,
       volume = {42},
          eid = {e089},
        pages = {e089},
          doi = {10.1017/pasa.2025.10060},
archivePrefix = {arXiv},
       eprint = {2506.13459},
 primaryClass = {astro-ph.EP},
       adsurl = {https://ui.adsabs.harvard.edu/abs/2025PASA...42...89B}
}

@article{Wandia_2025,
  title = {Upper Limits on Radio Emission from the K2-18 System},
  ISSN = {1365-2966},
  url = {http://dx.doi.org/10.1093/mnras/staf1998},
  DOI = {10.1093/mnras/staf1998},
  journal = {Monthly Notices of the Royal Astronomical Society},
  publisher = {Oxford University Press (OUP)},
  author = {Wandia,  Kelvin and Tremblay,  Chenoa and Garrett,  Michael A and Andersson,  Alex and Li,  Megan G and Gajjar,  Vishal and Beswick,  Robert J and Radcliffe,  Jack F and DeBoer,  David R and Demorest,  P B and Czech,  Daniel and Farah,  Wael and Heywood,  Ian and Siemion,  Andrew},
  year = {2025},
  month = nov 
}

@ARTICLE{Montet_2015,
       author = {{Montet}, Benjamin T. and {Morton}, Timothy D. and {Foreman-Mackey}, Daniel and {Johnson}, John Asher and {Hogg}, David W. and {Bowler}, Brendan P. and {Latham}, David W. and {Bieryla}, Allyson and {Mann}, Andrew W.},
        title = "{Stellar and Planetary Properties of K2 Campaign 1 Candidates and Validation of 17 Planets, Including a Planet Receiving Earth-like Insolation}",
      journal = {\apj},
         year = 2015,
        month = aug,
       volume = {809},
       number = {1},
          eid = {25},
        pages = {25},
          doi = {10.1088/0004-637X/809/1/25},
archivePrefix = {arXiv},
       eprint = {1503.07866},
 primaryClass = {astro-ph.EP},
       adsurl = {https://ui.adsabs.harvard.edu/abs/2015ApJ...809...25M}
}

@ARTICLE{Wright_2022,
       author = {{Wright}, Jason T. and {Haqq-Misra}, Jacob and {Frank}, Adam and {Kopparapu}, Ravi and {Lingam}, Manasvi and {Sheikh}, Sofia Z.},
        title = "{The Case for Technosignatures: Why They May Be Abundant, Long-lived, Highly Detectable, and Unambiguous}",
      journal = {\apjl},
         year = 2022,
        month = mar,
       volume = {927},
       number = {2},
          eid = {L30},
        pages = {L30},
          doi = {10.3847/2041-8213/ac5824},
archivePrefix = {arXiv},
       eprint = {2203.10899},
 primaryClass = {astro-ph.EP},
       adsurl = {https://ui.adsabs.harvard.edu/abs/2022ApJ...927L..30W}
}

@ARTICLE{Madhusudhan_2023,
       author = {{Madhusudhan}, Nikku and {Sarkar}, Subhajit and {Constantinou}, Savvas and {Holmberg}, M{\r{a}}ns and {Piette}, Anjali A.~A. and {Moses}, Julianne I.},
        title = "{Carbon-bearing Molecules in a Possible Hycean Atmosphere}",
      journal = {\apjl},
         year = 2023,
        month = oct,
       volume = {956},
       number = {1},
          eid = {L13},
        pages = {L13},
          doi = {10.3847/2041-8213/acf577},
archivePrefix = {arXiv},
       eprint = {2309.05566},
 primaryClass = {astro-ph.EP},
       adsurl = {https://ui.adsabs.harvard.edu/abs/2023ApJ...956L..13M}
}

@ARTICLE{Barclay_2021,
       author = {{Barclay}, Thomas and {Kostov}, Veselin B. and {Col{\'o}n}, Knicole D. and {Quintana}, Elisa V. and {Schlieder}, Joshua E. and {Louie}, Dana R. and {Gilbert}, Emily A. and {Mullally}, Susan E.},
        title = "{Stellar Surface Inhomogeneities as a Potential Source of the Atmospheric Signal Detected in the K2-18b Transmission Spectrum}",
      journal = {\aj},
         year = 2021,
        month = dec,
       volume = {162},
       number = {6},
          eid = {300},
        pages = {300},
          doi = {10.3847/1538-3881/ac2824},
archivePrefix = {arXiv},
       eprint = {2109.14608},
 primaryClass = {astro-ph.EP},
       adsurl = {https://ui.adsabs.harvard.edu/abs/2021AJ....162..300B}
}

@ARTICLE{Howard_2025,
       author = {{Howard}, Saburo and {Helled}, Ravit and {Bergermann}, Armin and {Redmer}, Ronald},
        title = "{The Possibility of Hydrogen-Water Demixing in Uranus, Neptune, K2-18b and TOI-270d}",
      journal = {arXiv e-prints},
         year = 2025,
        month = jul,
          eid = {arXiv:2507.06288},
        pages = {arXiv:2507.06288},
          doi = {10.48550/arXiv.2507.06288},
archivePrefix = {arXiv},
       eprint = {2507.06288},
 primaryClass = {astro-ph.EP},
       adsurl = {https://ui.adsabs.harvard.edu/abs/2025arXiv250706288H}
}

@ARTICLE{K218b_transit,
       author = {{Benneke}, Bj{\"o}rn and {Wong}, Ian and {Piaulet}, Caroline and {Knutson}, Heather A. and {Lothringer}, Joshua and {Morley}, Caroline V. and {Crossfield}, Ian J.~M. and {Gao}, Peter and {Greene}, Thomas P. and {Dressing}, Courtney and {Dragomir}, Diana and {Howard}, Andrew W. and {McCullough}, Peter R. and {Kempton}, Eliza M. -R. and {Fortney}, Jonathan J. and {Fraine}, Jonathan},
        title = "{Water Vapor and Clouds on the Habitable-zone Sub-Neptune Exoplanet K2-18b}",
      journal = {\apjl},
         year = 2019,
        month = dec,
       volume = {887},
       number = {1},
          eid = {L14},
        pages = {L14},
          doi = {10.3847/2041-8213/ab59dc},
archivePrefix = {arXiv},
       eprint = {1909.04642},
 primaryClass = {astro-ph.EP},
       adsurl = {https://ui.adsabs.harvard.edu/abs/2019ApJ...887L..14B}
}

@ARTICLE{Saide_2023,
       author = {{Saide}, Ramiro C. and {Farah}, Wael and {Sheikh}, Sofia Z. and {Pollak}, Alexander W. and {Siemion}, Andrew P.~V. and {Cruz}, Luigi F. and {Davis}, Roy H. and {DeBoer}, David R. and {Gajjar}, Vishal and {Karn}, Phil and {Masters}, Mark and {Shumaker}, Carol and {Singh}, Gurmehar and {Snodgrass}, Michael},
        title = "{Hycean Exoplanets as Targets for Technosignature Detection: A Case Study of K2-18 b in the 3-10 GHz Band}",
      journal = {Research Notes of the American Astronomical Society},
         year = 2023,
        month = nov,
       volume = {7},
       number = {11},
          eid = {233},
        pages = {233},
          doi = {10.3847/2515-5172/ad08b2},
       adsurl = {https://ui.adsabs.harvard.edu/abs/2023RNAAS...7..233S}
}

@ARTICLE{Tremblay_VLASS,
       author = {{Tremblay}, Chenoa D. and {Sofair}, Jared and {Steffes}, Lucy and {Myburgh}, Talon and {Czech}, Daniel and {Demorest}, Paul B. and {Donnachie}, Ross A. and {Pollak}, Alex W. and {Ruzindana}, Mark and {Siemion}, Andrew P.~V. and {Varghese}, Savin S. and {Sheikh}, Sofia},
        title = "{COSMIC's Large-Scale Search for Technosignatures during the VLA sky Survey: Survey Description and First Results}",
      journal = {arXiv e-prints},
         year = 2025,
        month = jan,
          eid = {arXiv:2501.17997},
        pages = {arXiv:2501.17997},
          doi = {10.48550/arXiv.2501.17997},
archivePrefix = {arXiv},
       eprint = {2501.17997},
 primaryClass = {astro-ph.IM},
       adsurl = {https://ui.adsabs.harvard.edu/abs/2025arXiv250117997T}
}

@ARTICLE{Choza_2023,
       author = {{Choza}, Carmen and {Bautista}, Daniel and {Croft}, Steve and {Siemion}, Andrew P.~V. and {Brzycki}, Bryan and {Bhattaram}, Krishnakumar and {Czech}, Daniel and {de Pater}, Imke and {Gajjar}, Vishal and {Isaacson}, Howard and {Lacker}, Kevin and {Lacki}, Brian and {Lebofsky}, Matthew and {MacMahon}, David H.~E. and {Price}, Danny and {Schoultz}, Sarah and {Sheikh}, Sofia and {Varghese}, Savin Shynu and {Morgan}, Lawrence and {Drew}, Jamie and {Worden}, S. Pete},
        title = "{The Breakthrough Listen Search for Intelligent Life: Technosignature Search of 97 Nearby Galaxies}",
      journal = {\aj},
         year = 2024,
        month = jan,
       volume = {167},
       number = {1},
          eid = {10},
        pages = {10},
          doi = {10.3847/1538-3881/acf576},
archivePrefix = {arXiv},
       eprint = {2312.03943},
 primaryClass = {astro-ph.IM},
       adsurl = {https://ui.adsabs.harvard.edu/abs/2024AJ....167...10C}
}

@ARTICLE{Li_2023,
       author = {{Li}, Megan G. and {Sheikh}, Sofia Z. and {Gilbertson}, Christian and {He}, Matthias Y. and {Isaacson}, Howard and {Croft}, Steve and {Sneed}, Evan L.},
        title = "{Developing a Drift Rate Distribution for Technosignature Searches of Exoplanets}",
      journal = {\aj},
         year = 2023,
        month = nov,
       volume = {166},
       number = {5},
          eid = {182},
        pages = {182},
          doi = {10.3847/1538-3881/acf83d},
archivePrefix = {arXiv},
       eprint = {2311.01427},
 primaryClass = {astro-ph.EP},
       adsurl = {https://ui.adsabs.harvard.edu/abs/2023AJ....166..182L}
}

@ARTICLE{Li_2022,
       author = {{Li}, Jian-Kang and {Zhao}, Hai-Chen and {Tao}, Zhen-Zhao and {Zhang}, Tong-Jie and {Xiao-Hui}, Sun},
        title = "{Drift Rates of Narrowband Signals in Long-term SETI Observations for Exoplanets}",
      journal = {\apj},
         year = 2022,
        month = oct,
       volume = {938},
       number = {1},
          eid = {1},
        pages = {1},
          doi = {10.3847/1538-4357/ac90bd},
archivePrefix = {arXiv},
       eprint = {2208.02511},
 primaryClass = {astro-ph.IM},
       adsurl = {https://ui.adsabs.harvard.edu/abs/2022ApJ...938....1L}
}

@ARTICLE{tremblay_cosmic,
       author = {{Tremblay}, C.~D. and {Varghese}, S.~S. and {Hickish}, J. and {Demorest}, P.~B. and {Ng}, C. and {Siemion}, A.~P.~V. and {Czech}, D. and {Donnachie}, R.~A. and {Farah}, W. and {Gajjar}, V. and {Lebofsky}, M. and {MacMahon}, D.~H.~E. and {Myburgh}, T. and {Ruzindana}, M. and {Bright}, J.~S. and {Erickson}, A. and {Lacker}, K.},
        title = "{COSMIC: An Ethernet-based Commensal, Multimode Digital Backend on the Karl G. Jansky Very Large Array for the Search for Extraterrestrial Intelligence}",
      journal = {\aj},
         year = 2024,
        month = jan,
       volume = {167},
       number = {1},
          eid = {35},
        pages = {35},
          doi = {10.3847/1538-3881/ad0fe0},
archivePrefix = {arXiv},
       eprint = {2310.09414},
 primaryClass = {astro-ph.IM},
       adsurl = {https://ui.adsabs.harvard.edu/abs/2024AJ....167...35T}
}

@techreport{perley_widar,
    author = {{Perley}, R.},
    title = "{Wide-Field Imaging with the EVLA: WIDAR Correlator Modes and Output Data Rates}",
    institution = "{National Radio Astronomy Observatory (NRAO)}",
    year = 2004,
    url="{https://library.nrao.edu/public/memos/evla/legacy/evlamemo64.pdf}"
}

@ARTICLE{Astropy,
       author = {{Astropy Collaboration} and {Robitaille}, Thomas P. and
         {Tollerud}, Erik J. and {Greenfield}, Perry and {Droettboom}, Michael and
         {Bray}, Erik and others},
        title = "{Astropy: A community Python package for astronomy}",
      journal = {\aap},
         year = 2013,
        month = oct,
       volume = {558},
          eid = {A33},
        pages = {A33},
          doi = {10.1051/0004-6361/201322068},
archivePrefix = {arXiv},
       eprint = {1307.6212},
 primaryClass = {astro-ph.IM},
       adsurl = {https://ui.adsabs.harvard.edu/abs/2013A&A...558A..33A}
}

@ARTICLE{Czech_2021,
       author = {{Czech}, Daniel and {Isaacson}, Howard and {Pearce}, Logan and {Cox}, Tyler and {Sheikh}, Sofia Z. and {Brzycki}, Bryan and {Buchner}, Sarah and {Croft}, Steve and {DeBoer}, David and {DeMarines}, Julia and {Drew}, Jamie and {Gajjar}, Vishal and {Lacki}, Brian C. and {Lebofsky}, Matt and {MacMahon}, David H.~E. and {Ng}, Cherry and {de Pater}, Imke and {Price}, Danny C. and {Siemion}, Andrew P.~V. and {Van Rooyen}, Ruby and {Pete Worden}, S.},
        title = "{The Breakthrough Listen Search for Intelligent Life: MeerKAT Target Selection}",
      journal = {\pasp},
         year = 2021,
        month = jun,
       volume = {133},
       number = {1024},
          eid = {064502},
        pages = {064502},
          doi = {10.1088/1538-3873/abf329},
archivePrefix = {arXiv},
       eprint = {2103.16250},
 primaryClass = {astro-ph.IM},
       adsurl = {https://ui.adsabs.harvard.edu/abs/2021PASP..133f4502C}
}

@ARTICLE{Siemion_2013,
       author = {{Siemion}, Andrew P.~V. and {Demorest}, Paul and {Korpela}, Eric and {Maddalena}, Ron J. and {Werthimer}, Dan and {Cobb}, Jeff and {Howard}, Andrew W. and {Langston}, Glen and {Lebofsky}, Matt and {Marcy}, Geoffrey W. and {Tarter}, Jill},
        title = "{A 1.1-1.9 GHz SETI Survey of the Kepler Field. I. A Search for Narrow-band Emission from Select Targets}",
      journal = {\apj},
         year = 2013,
        month = apr,
       volume = {767},
       number = {1},
          eid = {94},
        pages = {94},
          doi = {10.1088/0004-637X/767/1/94},
archivePrefix = {arXiv},
       eprint = {1302.0845},
 primaryClass = {astro-ph.GA},
       adsurl = {https://ui.adsabs.harvard.edu/abs/2013ApJ...767...94S}
}

@ARTICLE{Cocconi_1959,
       author = {{Cocconi}, Giuseppe and {Morrison}, Philip},
        title = "{Searching for Interstellar Communications}",
      journal = {\nat},
         year = 1959,
        month = sep,
       volume = {184},
       number = {4690},
        pages = {844-846},
          doi = {10.1038/184844a0},
       adsurl = {https://ui.adsabs.harvard.edu/abs/1959Natur.184..844C}
}

@INPROCEEDINGS{COSMIC,
       author = {{Hickish}, Jack and {Beasley}, Tony and {Bower}, Geoff and {Burke-Spolaor}, Sarah and {Croft}, Steve and {DeBoer}, Dave and {Demorest}, Paul and {Diamond}, Bill and {Gajjar}, Vishal and {Law}, Casey and {Lazio}, Joseph and {Manley}, Jason and {Paragi}, Zsolt and {Ransom}, Scott and {Siemion}, Andrew},
        title = "{Commensal, Multi-user Observations with an Ethernet-based Jansky Very Large Array}",
    booktitle = {Bulletin of the American Astronomical Society},
         year = 2019,
       volume = {51},
        month = sep,
          eid = {269},
        pages = {269},
       adsurl = {https://ui.adsabs.harvard.edu/abs/2019BAAS...51g.269H}
}

@ARTICLE{Sarkis2018,
       author = {{Sarkis}, Paula and {Henning}, Thomas and {K{\"u}rster}, Martin and {Trifonov}, Trifon and {Zechmeister}, Mathias and {Tal-Or}, Lev and {Anglada-Escud{\'e}}, Guillem and {Hatzes}, Artie P. and {Lafarga}, Marina and {Dreizler}, Stefan and {Ribas}, Ignasi and {Caballero}, Jos{\'e} A. and {Reiners}, Ansgar and {Mallonn}, Matthias and {Morales}, Juan C. and {Kaminski}, Adrian and {Aceituno}, Jes{\'u}s and {Amado}, Pedro J. and {B{\'e}jar}, Victor J.~S. and {Hagen}, Hans-J{\"u}rgen and {Jeffers}, Sandra and {Quirrenbach}, Andreas and {Launhardt}, Ralf and {Marvin}, Christopher and {Montes}, David},
        title = "{The CARMENES Search for Exoplanets around M Dwarfs: A Low-mass Planet in the Temperate Zone of the Nearby K2-18}",
      journal = {\aj},
         year = 2018,
        month = jun,
       volume = {155},
       number = {6},
          eid = {257},
        pages = {257},
          doi = {10.3847/1538-3881/aac108},
archivePrefix = {arXiv},
       eprint = {1805.00830},
 primaryClass = {astro-ph.EP},
       adsurl = {https://ui.adsabs.harvard.edu/abs/2018AJ....155..257S}
}

@ARTICLE{2018AJ....156..123A,
       author = {{Astropy Collaboration} and {Price-Whelan}, A.~M. and {Sip{\H{o}}cz}, B.~M. and {G{\"u}nther}, H.~M. and {Lim}, P.~L. and {Crawford}, S.~M. and {Conseil}, S. and {Shupe}, D.~L. and {Craig}, M.~W. and {Dencheva}, N. and {Ginsburg}, A. and {VanderPlas}, J.~T. and {Bradley}, L.~D. and {P{\'e}rez-Su{\'a}rez}, D. and {de Val-Borro}, M. and {Aldcroft}, T.~L. and {Cruz}, K.~L. and {Robitaille}, T.~P. and {Tollerud}, E.~J. and {Ardelean}, C. and {Babej}, T. and {Bach}, Y.~P. and {Bachetti}, M. and {Bakanov}, A.~V. and {Bamford}, S.~P. and {Barentsen}, G. and {Barmby}, P. and {Baumbach}, A. and {Berry}, K.~L. and {Biscani}, F. and {Boquien}, M. and {Bostroem}, K.~A. and {Bouma}, L.~G. and {Brammer}, G.~B. and {Bray}, E.~M. and {Breytenbach}, H. and {Buddelmeijer}, H. and {Burke}, D.~J. and {Calderone}, G. and {Cano Rodr{\'\i}guez}, J.~L. and {Cara}, M. and {Cardoso}, J.~V.~M. and {Cheedella}, S. and {Copin}, Y. and {Corrales}, L. and {Crichton}, D. and {D'Avella}, D. and {Deil}, C. and {Depagne}, {\'E}. and {Dietrich}, J.~P. and {Donath}, A. and {Droettboom}, M. and {Earl}, N. and {Erben}, T. and {Fabbro}, S. and {Ferreira}, L.~A. and {Finethy}, T. and {Fox}, R.~T. and {Garrison}, L.~H. and {Gibbons}, S.~L.~J. and {Goldstein}, D.~A. and {Gommers}, R. and {Greco}, J.~P. and {Greenfield}, P. and {Groener}, A.~M. and {Grollier}, F. and {Hagen}, A. and {Hirst}, P. and {Homeier}, D. and {Horton}, A.~J. and {Hosseinzadeh}, G. and {Hu}, L. and {Hunkeler}, J.~S. and {Ivezi{\'c}}, {\v{Z}}. and {Jain}, A. and {Jenness}, T. and {Kanarek}, G. and {Kendrew}, S. and {Kern}, N.~S. and {Kerzendorf}, W.~E. and {Khvalko}, A. and {King}, J. and {Kirkby}, D. and {Kulkarni}, A.~M. and {Kumar}, A. and {Lee}, A. and {Lenz}, D. and {Littlefair}, S.~P. and {Ma}, Z. and {Macleod}, D.~M. and {Mastropietro}, M. and {McCully}, C. and {Montagnac}, S. and {Morris}, B.~M. and {Mueller}, M. and {Mumford}, S.~J. and {Muna}, D. and {Murphy}, N.~A. and {Nelson}, S. and {Nguyen}, G.~H. and {Ninan}, J.~P. and {N{\"o}the}, M. and {Ogaz}, S. and {Oh}, S. and {Parejko}, J.~K. and {Parley}, N. and {Pascual}, S. and {Patil}, R. and {Patil}, A.~A. and {Plunkett}, A.~L. and {Prochaska}, J.~X. and {Rastogi}, T. and {Reddy Janga}, V. and {Sabater}, J. and {Sakurikar}, P. and {Seifert}, M. and {Sherbert}, L.~E. and {Sherwood-Taylor}, H. and {Shih}, A.~Y. and {Sick}, J. and {Silbiger}, M.~T. and {Singanamalla}, S. and {Singer}, L.~P. and {Sladen}, P.~H. and {Sooley}, K.~A. and {Sornarajah}, S. and {Streicher}, O. and {Teuben}, P. and {Thomas}, S.~W. and {Tremblay}, G.~R. and {Turner}, J.~E.~H. and {Terr{\'o}n}, V. and {van Kerkwijk}, M.~H. and {de la Vega}, A. and {Watkins}, L.~L. and {Weaver}, B.~A. and {Whitmore}, J.~B. and {Woillez}, J. and {Zabalza}, V. and {Astropy Contributors}},
        title = "{The Astropy Project: Building an Open-science Project and Status of the v2.0 Core Package}",
      journal = {\aj},
         year = 2018,
        month = sep,
       volume = {156},
       number = {3},
          eid = {123},
        pages = {123},
          doi = {10.3847/1538-3881/aabc4f},
archivePrefix = {arXiv},
       eprint = {1801.02634},
 primaryClass = {astro-ph.IM},
       adsurl = {https://ui.adsabs.harvard.edu/abs/2018AJ....156..123A}
}

@ARTICLE{2013A&A...558A..33A,
       author = {{Astropy Collaboration} and {Robitaille}, Thomas P. and
         {Tollerud}, Erik J. and {Greenfield}, Perry and {Droettboom}, Michael and
         {Bray}, Erik and {Aldcroft}, Tom and {Davis}, Matt and
         {Ginsburg}, Adam and {Price-Whelan}, Adrian M. and
         {Kerzendorf}, Wolfgang E. and {Conley}, Alexander and {Crighton}, Neil and
         {Barbary}, Kyle and {Muna}, Demitri and {Ferguson}, Henry and
         {Grollier}, Fr{\'e}d{\'e}ric and {Parikh}, Madhura M. and
         {Nair}, Prasanth H. and {Unther}, Hans M. and {Deil}, Christoph and
         {Woillez}, Julien and {Conseil}, Simon and {Kramer}, Roban and
         {Turner}, James E.~H. and {Singer}, Leo and {Fox}, Ryan and
         {Weaver}, Benjamin A. and {Zabalza}, Victor and {Edwards}, Zachary I. and
         {Azalee Bostroem}, K. and {Burke}, D.~J. and {Casey}, Andrew R. and
         {Crawford}, Steven M. and {Dencheva}, Nadia and {Ely}, Justin and
         {Jenness}, Tim and {Labrie}, Kathleen and {Lim}, Pey Lian and
         {Pierfederici}, Francesco and {Pontzen}, Andrew and {Ptak}, Andy and
         {Refsdal}, Brian and {Servillat}, Mathieu and {Streicher}, Ole},
        title = "{Astropy: A community Python package for astronomy}",
      journal = {\aap},
         year = "2013",
        month = "Oct",
       volume = {558},
          eid = {A33},
        pages = {A33},
          doi = {10.1051/0004-6361/201322068},
archivePrefix = {arXiv},
       eprint = {1307.6212},
 primaryClass = {astro-ph.IM},
       adsurl = {https://ui.adsabs.harvard.edu/abs/2013A&A...558A..33A}
}

@article{wandia2025upper,
  title={Upper Limits on Radio Emission from the K2-18 System},
  author={Wandia, Kelvin and Tremblay, Chenoa and Garrett, Michael A and Andersson, Alex and Li, Megan G and Gajjar, Vishal and Beswick, Robert J and Radcliffe, Jack F and DeBoer, David R and Demorest, PB and others},
  journal={Monthly Notices of the Royal Astronomical Society},
  pages={staf1998},
  year={2025},
  publisher={Oxford University Press}
}
\bibliographystyle{aasjournal}



\end{document}